\documentclass[11pt,xcolor=dvipsnames,fleqn]{article}
\DeclareMathAlphabet{\scr}{U}{rsfs}{m}{n}

\pdfoutput=1
\usepackage{scalerel,stackengine}
\usepackage{subfigure}
\usepackage{latexsym}
\usepackage{epsfig}
\usepackage[mathscr]{eucal}
\usepackage{amsfonts}
\usepackage{amscd}
\usepackage{multirow}
\usepackage{cite}
\usepackage{array}
\usepackage{amssymb}
\usepackage{amsthm}
\usepackage{colordvi}
\usepackage[centertags]{amsmath}
\usepackage{indentfirst}
\usepackage{geometry}
\usepackage{float} 
\usepackage{enumerate}
\usepackage{graphicx}
\usepackage{booktabs}
\usepackage{theorem}
\usepackage[footnotesize]{caption}
\usepackage{soul}
\usepackage{mcite}
\usepackage{slashed}%slash notation for gamma matrices
\usepackage{braket}%for bra and ket notations
\usepackage{color,xcolor}
\usepackage{bbm}

\usepackage[utf8]{inputenc}
\usepackage{fancyvrb}% provides improved Verbatim command
\usepackage{framed}
\usepackage{xspace}
\usepackage{todonotes}
\usepackage{siunitx}%Allows for formatting units via \SI{number}{Unit}

\usepackage{psfrag,rotating}
\usepackage{epstopdf}
\usepackage[normalem]{ulem} % for text strikeout
\usepackage{cancel} 

\numberwithin{figure}{section}
\numberwithin{table}{section}

\allowdisplaybreaks

\usepackage{cleveref}
\crefname{chapter}{Chapter}{Chapter}
\crefname{section}{Sec.}{Secs.}
\crefname{table}{Tab.}{Tabs.}
\crefname{figure}{Fig.}{Figs.}
\crefname{equation}{Eq.}{Eqs.}
\crefname{appendix}{Appendix\ }{Appendix\ }

\begin{document}

	\title{

		\vspace*{-3.7cm}
		\vspace*{2.7cm}
		\textbf{One-loop QED and Weak Corrections to  $\gamma \gamma \to H^\pm H^\mp$ in the Inert Doublet Model \\[4mm]}}

	\date{}
	\author{
		Hamza Abouabid$^{1\,}$\footnote{E-mail:
			\texttt{hamza.abouabid@gmail.com}} ,
		Abdesslam Arhrib$^{2,3\,}$\footnote{E-mail:
			\texttt{aarhrib@uae.ac.ma}} ,
		Jaouad El Falaki$^{4\,}$\footnote{E-mail: \texttt{jaouad.elfalaki@gmail.com}},
		Bin Gong$^{5,6\,}$\footnote{{E-mail: \texttt{twain@ihep.ac.cn}}},
		\\
	%	Wenhai Xie$^{3,4\,}$\footnote{E-mail:
	%		\texttt{xiewh@ihep.ac.cn}},
		Qi-Shu Yan$^{6,7\,}$\footnote{E-mail: \texttt{yanqishu@ucas.ac.cn}},
		\\[5mm]
			{\small\it $^1$ LPRI, Moroccan School of Engineering Sciences (EMSI), Casablanca 20250, Morocco.} \\[3mm]
		{\small\it
			$^2$Universit\'{e} Abdelmalek Essaadi, FSTT, B. 416, Tangier, Morocco.} \\[3mm]
			{\small\it
			$^3$LAPTh, CNRS, Universit\'e Savoie Mont-Blanc, 9 Chemin de Bellevue, 74940, Annecy, France.} \\[3mm]
		{\small\it $^4$ LPTHE, Physics Department, Faculty of Sciences, Ibnou Zohr University, P.O.B. 8106 Agadir, Morocco.} \\[3mm]
%		{\small\it $^5$ Theory Division, Institute of High Energy Physics, Chinese Academy of Sciences, 19B Yuquan Road, Shijingshan District,Beijing 100049, China.} \\[3mm]	
		{\small\it $^5$ Theory Division, Institute of High Energy Physics, Chinese Academy of Sciences},\\
{\small\it 19B Yuquan Road, Shijingshan District, Beijing 100049, China.} \\[3mm]	
%		{\small\it $^6$School of Physics Sciences, University of Chinese Academy of Sciences, Beijing 100049, China.} \\[3mm]
		{\small\it $^6$University of Chinese Academy of Sciences},\\
{\small\it No.1 Yanqihu East Rd, Huairou District, Beijing 101408, PR China.}\\[3mm]
%		{\small\it $^7$ Center for Future High Energy Physics, Chinese Academy of Sciences,  Beijing 100049, China.} \\[3mm]
		{\small\it $^7$ Center for Future High Energy Physics, Chinese Academy of Sciences},\\
{\small\it19B Yuquan Road, Shijingshan District, Beijing 100049, China.} \\[3mm]
	}
	\maketitle
	
\thispagestyle{empty}
%\vspace*{2.5cm}
%\vspace{0.5cm}
%\begin{center}
\vspace*{0.1cm}

\begin{abstract}
    \noindent
We present a complete one-loop analysis of charged scalar boson pair production in photon-photon collisions, $\gamma\gamma \to H^\pm H^\mp$, within the framework of the Inert Doublet Model (IDM). The calculation is carried out in the on-shell renormalization scheme and incorporates both weak corrections and QED effects, including soft and hard photon radiation. Virtual loop contributions and real emission processes are computed using the Feynman diagrammatic method, ensuring the cancellation of ultraviolet and infrared divergences. To properly account for the Coulomb singularity that arises in the QED sector near threshold, we introduce the resummed cross section based on the Sommerfeld factor.  The IDM parameter space is explored under theoretical consistency conditions, collider limits, and dark matter constraints, and three representative scenarios are studied in detail. We find that the magnitude of the quantum corrections is strongly controlled by the absolute value of the trilinear scalar coupling $\lambda_{h^0 H^+ H^-}$, which correlates with the charged scalar mass. When all constraints are applied, the weak corrections are typically in the range of $-12\%$ to $-7\%$ at $\sqrt{s}=250$~GeV, and between $-15\%$ and $6\%$ at $\sqrt{s}=500$~GeV. At higher energies, such as $\sqrt{s}=1$~TeV, the corrections can become very large, ranging from about $-20\%$ up to $+60\%$.  
Our findings highlight the significant role of higher-order effects in photon-photon collisions and establish $\gamma\gamma \to H^\pm H^\mp$ as a promising process to investigate the charged scalar sector of the IDM at future high-energy photon colliders. Several benchmark points are proposed to facilitate future experimental searches.  
\end{abstract}
\thispagestyle{empty}
\def\thefootnote{\arabic{footnote}}
\setcounter{footnote}{0}
\newpage
\setcounter{page}{1}

\section{Introduction}
\label{sec:introduction}
%%%%%%%%%%%%%%%%%%%%%%%%

The Large Hadron Collider (LHC) has achieved remarkable milestones through its successful two runs at $7 \oplus 8$ TeV and $13$ TeV. 
These runs culminated in the discovery of a new $125$ GeV spinless scalar boson, whose properties are consistent with the Standard Model (SM) Higgs boson, 
as confirmed by the ATLAS and CMS collaborations in 2012 \cite{Aad:2012tfa, Chatrchyan:2012xdj}. 
This discovery, including subsequent observations of the Higgs boson produced in association with a pair of top and anti-top quarks $h^0t\bar{t}$~\cite{Aaboud:2018urx, Sirunyan:2018hoz}, its decay into bottom and anti-bottom quarks $h^0 \to b \bar{b}$~\cite{Aaboud:2018zhk,Sirunyan:2018kst} and also an evidence for the observation of 
$h^0\to \gamma Z^0$ \cite{ATLAS:2023yqk} which agrees with the theoretical expectation within 1.9 standard deviations, has provided robust evidence supporting the SM as an effective field theory at the electroweak scale of a more fundamental theory yet to be discovered.

One of the primary objectives for the current LHC run at 14 TeV, including the future High Luminosity (HL-LHC) option, is to improve the current measurements and reduce uncertainties to a few percent \cite{Dawson:2013bba, Zeppenfeld:2000td, Gianotti:2000tz, Cepeda:2019klc}. Additionally, the future LHC run aims to achieve new measurements, such as the Higgs trilinear coupling and its decays into $\gamma Z^0$ and $\mu^+\mu^-$. 

As we enter an era of precision Higgs physics that will improve with the current LHC run, the need for accurate theoretical predictions becomes paramount. Future lepton colliders~\cite{Moortgat-Picka:2015yla, Fujii:2015jha, Fujii:2017vwa}, such as the International Linear Collider (ILC)\cite{Moortgat-Picka:2015yla}, the Future Circular Collider (FCC-ee) \cite{Gomez-Ceballos:2013zzn}, or the Compact Linear Collider (CLIC)\cite{Battaglia:2004mw, Aicheler:2012bya, Linssen:2012hp}, promise unprecedented precision in 
measuring Higgs properties. These facilities also offer the possibility of photon–photon collisions through Compton backscattering, 
	opening up new avenues for precision Higgs studies, 
	as recently demonstrated in analyses of Higgs-pair production at photon colliders~\cite{Berger:2025ijd}, 
	which further highlight their potential for probing extended Higgs sectors. In particular, photon colliders provide a clean environment to study charged scalar bosons, since they directly couple to photons through their electric charge, without the need for additional mediators.
Related photon-induced production mechanisms have also been explored at the LHC in the context of charged Higgs bosons~\cite{Arroyo-Urena:2025boh}.

Despite these great achievements made by the LHC program, the SM remains incomplete, as it fails to address several fundamental questions, 
such as the nature of dark matter, neutrino masses, and the matter-antimatter asymmetry. To explore these mysteries, extensions of the SM, like the IDM, offer promising solutions by introducing additional scalar particles.  

The IDM extends the SM by introducing an additional scalar doublet, denoted as $H_2$ \cite{Deshpande:1977rw,Ma:2006km,Barbieri:2006dq}. This doublet complements the SM Higgs doublet, $H_1$, which is responsible for electroweak symmetry breaking (EWSB). The new doublet $H_2$, in contrast, does not partake in the EWSB process and remains inert.
Moreover, the model is protected by a discrete $\mathbb{Z}_2$ symmetry under which $H_2$ is odd and all the other fields are even.
Such $\mathbb{Z}_2$ symmetry prevents direct couplings of $H_2$ to SM fermions and ensures the stability of the lightest odd particle (LOP), which can serve as a dark matter candidate. Although this concept was first put forth more than 40 years ago \cite{Deshpande:1977rw}, it wasn't until 2006 that there was renewed
phenomenological interest in this model \cite{Gustafsson:2007pc, Hambye:2007vf, Agrawal:2008xz, Dolle:2009fn, Andreas:2009hj, Barbieri:2006dq}.  

After EWSB takes place, the scalar spectrum of the IDM contains five scalar bosons: a CP-even Higgs $h^0$ associated with the observed 125 GeV scalar particle that completely mimics the SM Higgs; two neutral scalar bosons $H^0$ and $A^0$; and a pair of charged scalar bosons $H^\pm$. Numerous phenomenological studies have explored various aspects of the IDM, including its implications for dark matter, astrophysics, and collider constraints \cite{Kalinowski:2018ylg, Belyaev:2016lok, Arhrib:2013ela,Ilnicka:2015jba,Braathen:2024ckk}. These studies have consistently shown that the IDM remains in agreement with both theoretical predictions and experimental bounds.  
In particular, the charged scalar $H^\pm$
	plays a crucial role, as its mass is constrained by electroweak precision tests and dark matter relic density, while flavor observables do not impose direct bounds due to the absence of couplings to fermions in the IDM.

In the context of $e^+ e^-$ colliders, the charged scalar can be copiously produced pairwise via $e^+e^- \to H^\pm H^\mp$. In the $\gamma\gamma$ option of these colliders,
one can also access the inert charged scalars via $\gamma\gamma \to H^\pm H^\mp$. The process $\gamma\gamma \to H^\pm H^\mp$ emerges as a particularly interesting probe of the IDM. 
This production channel offers direct access to the charged scalar sector, complementing searches in electron-positron collisions ($e^+e^- \to H^\pm H^\mp$). 
At tree level, the process occurs via $t$- and $u$-channel exchanges of the charged scalar boson $H^\pm$, as well as a contact interaction. 
However, loop-level contributions introduce sensitivity to the neutral scalars and other model parameters.  Moreover, the one-loop contributions are ultraviolet (UV) divergent and infrared (IR) sensitive, requiring a consistent renormalization scheme and the inclusion of soft photon bremsstrahlung to obtain finite predictions.

While the process $e^+e^- \to H^\pm H^\mp$ has been extensively studied at the one-loop level in models such as the Minimal Supersymmetric Standard Model (MSSM), the Two Higgs Doublet Model (2HDM), and the IDM \cite{Komamiya:1988rs,Arhrib:1994xp,Arhrib:1998gr,Guasch:2001hk,Heinemeyer:2016wey,Moretti:2002sn,Hashemi:2013sja,Abouabid:2022rnd}, 
the $\gamma \gamma \to H^\pm H^\mp$ channel offers potential advantages at high energies due to the absence of $s$-channel suppression. 
The one-loop corrections for $\gamma \gamma \to H^\pm H^\mp$ in the framework of the 2HDM and the MSSM have been studied in~\cite{Wang:2005pjs,Sonmez:2018unx,Ma:1996nq,Demirci:2020zgt}, but a comprehensive evaluation of the full one-loop electroweak corrections in the IDM is still lacking. 

These kinds of calculations are crucial to match the expected precision at future lepton colliders and to allow for in-depth studies of charged scalar boson properties 
	if they are discovered in the future. Furthermore, the one-loop contribution is sensitive to various scalar potential parameters through neutral Higgs boson exchange,
	offering the possibility to distinguish between different models using precise measurements of the cross-section. 
	In addition, such processes provide complementary information to dark matter searches, since the charged scalar mass and couplings indirectly affect the relic density through coannihilation channels. 
	
	Moreover, radiative or higher-order corrections within the IDM have already been investigated in several works, 
	covering a range of observables such as Higgs decays, trilinear scalar couplings and scalar pair production processes\cite{Abouabid:2020eik, Arhrib:2015hoa,Kanemura:2016sos,Falaki:2023tyd, Aiko:2023nqj,He:2024bwh, Aiko:2021nkb, Ramsey-Musolf:2021ldh,Banerjee:2019luv,Banerjee:2021oxc,Banerjee:2021anv,Banerjee:2021xdp,Banerjee:2021hal,Banerjee:2016vrp,Kanemura:2019kjg,Braathen:2019zoh,Braathen:2019pxr, Senaha:2018xek,Bahl:2023eau}.
	These studies highlight the significant impact of loop effects
	for precision predictions in the IDM framework.

The structure of this paper is as follows: Section \ref{sec:model} provides a brief overview of the IDM and its theoretical and experimental constraints. 
Section \ref{sec:LO} presents both the leading-order (LO) and next-to-leading-order (NLO) calculations for the process $\gamma\gamma \rightarrow H^\pm H^\mp$. 
It introduces the relevant Feynman diagrams and the calculation of lowest-order cross-sections. 
The section also details the loop-level corrections, including electroweak and virtual photon contributions, and explains the renormalization scheme used. 
The importance of soft and hard photon emissions in canceling infrared (IR) divergences is examined to ensure accurate NLO results. 
Section \ref{sec:results} presents the one-loop contributions to the $\gamma \gamma \to H^\pm H^\mp$ process and some typical benchmark points. 
Finally, Section \ref{sec:conclusions} concludes the paper with a discussion of the results and future prospects.

%%%%%%%%%%%%%%%%%%%%%%%%%%%%%
\section{Model and constraints}
\label{sec:model}
\subsection{The IDM}
%%%%%%%%%%%%%%%%%%%%%%%%%%%%%
%%%%%%%%%%%%%%%%%%%%%%%%%%%%%
 One of the simplest SM extensions that can handle the dark matter feature is the IDM.
It consists of supplementing the SM Higgs doublet $H_1$ with an inert  doublet $H_2$.
Under the new discrete $\mathbb{Z}_2$ symmetry, the doublet $H_2$ is odd and does not couple with fermions.
$\mathbb{Z}_2$ prevents $H_2$ from developing a vacuum expectation value (VEV).
The scalar field representations in the IDM are formulated as:
\begin{eqnarray}
H_1 = \left(\begin{array}{c}
G^\pm \\
\frac{1}{\sqrt{2}}(v + h^0 + i G^0) \
\end{array} \right)
\qquad , \qquad
H_2 = \left( \begin{array}{c}
H^\pm\\
\frac{1}{\sqrt{2}}(H^0 + i A^0) \
\end{array} \right).
\end{eqnarray}

In this representation, $G^0$ and $G^\pm$ are the Goldstone bosons, which become the longitudinal components of the $Z^0$ and $W^\pm$ bosons post-EWSB. 
The parameter $v$ symbolizes VEV of the SM Higgs doublet $H_1$.

The general potential for the IDM, respecting renormalizability, CP conservation, and $\mathbb{Z}_2$ symmetry, is expressed as:
\begin{eqnarray}
V &=& \mu_1^2 |H_1|^2 + \mu_2^2 |H_2|^2 + \lambda_1 |H_1|^4 +
\lambda_2 |H_2|^4 + \lambda_3 |H_1|^2 |H_2|^2 + \lambda_4
|H_1^\dagger H_2|^2 \nonumber \\
&+&\frac{\lambda_5}{2} \left\{ (H_1^\dagger H_2)^2 + {\rm h.c} \right\}
\label{potential}
\end{eqnarray}
The $\mathbb{Z}_2$ symmetry notably prohibits mixing terms like $\mu_{12}^2 (H_1^\dagger H_2 + h.c.)$. Ensuring Hermiticity of the potential means that all $\lambda_i$ coefficients (i=1, ..., 4) are dimensionless and real, and the phase of $\lambda_5$ can be rotated away through a redefinition of $H_1$ and $H_2$.

After the EWSB, the physical scalar spectrum consists of the known SM Higgs boson $h^0$ with a mass of 125 GeV, along with four new scalar states: two charged  
($H^\pm$) and two neutral ($A^0$ and $H^0$) particles. Their masses are determined as follows:

\begin{eqnarray}
&& m_{h^0}^2 = - 2 \mu_1^2 = 2 \lambda_1 v^2 \nonumber \\
&& m_{H^{\pm}}^2 = \mu_2^2 + \frac{1}{2} \lambda_3 v^2 \nonumber \\
&& m_{A^0}^2 = \mu_2^2 + \lambda_S v^2 \nonumber \\
&& m_{H^0}^2 = \mu_2^2 + \lambda_L v^2
\label{spect.IDM}
\end{eqnarray}
with $\lambda_{S}$ and $\lambda_{L}$ being defined as:
\begin{eqnarray}
\lambda_{S} = \frac{1}{2} (\lambda_3 + \lambda_4 - \lambda_5)\quad \rm{and} \quad \lambda_{L} = \frac{1}{2} (\lambda_3 + \lambda_4 + \lambda_5)
\end{eqnarray}

Consequently, formulae for the $\lambda_i$ as a function of the masses can be found from the masses provided above:
\begin{eqnarray}
\lambda_1&=&\dfrac{m_{h^0}^2}{2v^2} \nonumber\\
\lambda_3&=&2\left(\dfrac{m_{H^\pm}^2-\mu_{2}^2}{v^2}\right) \nonumber\\
\lambda_4&=&\dfrac{m_{H^0}^2+m_{A^0}^2 -2 m_{H^\pm} ^2}{v^2} \nonumber\\
\lambda_5&=&\dfrac{m_{H^0}^2-m_{A^0}^2}{v^2}
\label{eq:lams}
\end{eqnarray}

Initially, there are eight independent parameters in the IDM: the VEV, $\lambda_{1,\dots,5}$, $\mu_{1}$, and $\mu_{2}$.
Using the minimization condition and 
recognizing the VEV as a fixed parameter determined by the $Z^0$ boson mass, fine-structure constant, and Fermi constant $G_F$, we reduce these parameters to six:

\begin{eqnarray}
\mathcal{P}=\left\{ \mu_2^2, m_{h^0}, m_{H^\pm}, m_{H^0}, m_{A^0}, \lambda_2 \right\}
\label{param.IDM}
\end{eqnarray}

The leading order triple scalar couplings that are essential to our analysis can be obtained from the scalar potential mentioned above as follows:

\begin{eqnarray}
&&h^0 H^\pm H^\mp = -v \lambda_3 \equiv v\lambda_{h^0H^\pm H^\mp}
%\equiv v\lambda_{h^0SS}
\nonumber\\
&&h^0 A^0 A^0 = -2v \lambda_S \equiv v\lambda_{h^0A^0 A^0}\nonumber\\
&&h^0 H^0 H^0 = -2v \lambda_L \equiv v\lambda_{h^0H^0 H^0}  \nonumber\\
&& H^\pm H^\mp H^\pm H^\mp = -4 \lambda_2  \nonumber\\
&& H^\pm H^\mp G^\pm G^\mp = -( \lambda_3  +\lambda_4  )  \nonumber\\
&& H^\pm H^\pm G^\mp G^\mp = -2 \lambda_5  \nonumber\\
&& H^\pm H^\mp A^0 A^0  =  H^\pm H^\mp H^0 H^0 =-2 \lambda_2    \nonumber\\
&& H^\pm H^\mp h^0 h^0   =  H^\pm H^\mp G^0 G^0 =  -\lambda_3  
\label{scalar-coup}
\end{eqnarray}

The relevant part of the Lagrangian describing the interaction  of the gauge bosons with the charged scalar boson is:
\begin{eqnarray}
{\cal L} &=& \frac{e}{2s_W}W_{\mu}^+ ( (H^- \stackrel{\leftrightarrow}{\partial}^{\mu} A^0)-
 i (H^- \stackrel{\leftrightarrow}{\partial}^{\mu} h^0)+
  i  (H^- \stackrel{\leftrightarrow}{\partial}^{\mu} H^0))+ h.c\nonumber\\
&&+ ( ie A_\mu+ i\frac{g (c_W^2-s_W^2)}{2 s_W c_W} Z_{\mu}^+)  (H^\mp\stackrel{\leftrightarrow}{\partial}^{\mu} H^\pm )  +\nonumber\\
 &&  ie^2 g^{\mu \nu} (  \frac{1 }{2 s_W^2 }  W_\mu W_\nu +  \frac{c_W^2-s_W^2 }{2 c_W^2 s_W^2 }   Z_\mu Z_\nu + 2 A_\mu A_\nu) H^+ H^-    \label{cov-der}
\end{eqnarray}
Where $e$ is the electric charge and $\{s,c\}_W$  is the sine and cosine of the Weinberg angle. 

\subsection{Theoretical and Experimental Framework Constraints}
\label{sec:thexcons}
This section delineates the constraints imposed on the IDM parameter space, classified into theoretical and experimental categories.

\subsubsection{Theoretical Constraints}
\label{sec:thecon}
The IDM's consistency with EWSB principles necessitates a series of constraints on its scalar sector.
\begin{itemize}
\item The perturbativity constraint is first considered, requiring all quartic scalar potential terms (Eq.~(\ref{potential})) to remain within the perturbative regime, 
thus mandating $|\lambda_i| \le 8 \pi $.

\item  Vacuum stability, ensuring the scalar potential is bounded from below at large field values, imposes specific conditions on the model parameters 
as per \cite{Deshpande:1977rw}:
\begin{equation}
\lambda_{1} > 0, \lambda_{2} > 0, \quad \lambda_3+2\sqrt{\lambda_1\lambda_2} > 0 \quad
\rm{and} \quad \lambda_3 + \lambda_4 -|\lambda_5| + 2\sqrt{\lambda_1 \lambda_2} >0.
\end{equation}

\item  The charge-conserving vacuum condition, as a sufficient criterion outlined in \cite{Ginzburg:2010wa}, is $\lambda_4-|\lambda_5|\leq 0$.

\item  The inert vacuum conditions, ensuring the CP-conserving minimum is global \cite{Ginzburg:2010wa}, require:
\begin{eqnarray}
m_{h^0}^2, m_{A^0}^2, m_{H^0}^2, m_{H^\pm}^2 >0 \qquad {\rm and} \qquad
v^2>-
\mu_2^2/\sqrt{\lambda_1\lambda_2}.
\label{eq:inertvac}
\end{eqnarray}

\item  Finally, to maintain unitarity at high-energy scalar boson scatterings, an upper limit of $8\pi$ is imposed on all eigenvalues determined via the method described in \cite{Lee:1977eg}:
\begin{eqnarray}
&&e_{1,2}=\lambda_3 \pm \lambda_4 \quad , \quad
e_{3,4}= \lambda_3 \pm \lambda_5,\\
&&e_{5,6}= \lambda_3+ 2 \lambda_4 \pm 3\lambda_5\quad , \quad
e_{7,8}=-\lambda_1 - \lambda_2 \pm \sqrt{(\lambda_1 - \lambda_2)^2 +\lambda_4^2},
\\
&&
e_{9,10}= -3\lambda_1 - 3\lambda_2 \pm \sqrt{9(\lambda_1 - \lambda_2)^2 + (2\lambda_3 +
\lambda_4)^2},
\\
&&
e_{11,12}=
-\lambda_1 - \lambda_2 \pm \sqrt{(\lambda_1 - \lambda_2)^2 + \lambda_5^2}.
\end{eqnarray}
\end{itemize}

\subsubsection{Experimental constraints}
\label{sec:excon}

The IDM parameters must satisfy a number of experimental requirements, in addition to the theoretical ones. These include constraints from Electroweak Precision Tests (EWPT), LEP and LHC searches, as well as dark matter observations.  

\begin{itemize}
	
	\item We apply indirect constraints from EWPT using the oblique parameters $S$, $T$, and $U$ \cite{Peskin:1991sw, Barbieri:2006dq, Tanabashi:2018oca}, 
	following the procedure of Ref.~\cite{Abouabid:2020eik}.  
	
	\item In the IDM, only the $H_1$ doublet couples to fermions, while $H_2$ does not couple to SM fermions due to the exact $\mathbb{Z}_2$ symmetry. 
	The charged scalar boson $H^\pm$ contributes only through loop corrections to the decays $h^0\to \gamma \gamma$ and $h^0\to \gamma Z^0$, 
	while the couplings of the SM-like Higgs $h^0$ to fermions and to $W^\pm$ and $Z^0$ bosons remain identical to the SM.  
	As a result, the main Higgs production mechanisms at the LHC (gluon fusion, vector boson fusion, Higgsstrahlung, and associated production with top-quark pairs) are unaffected and remain SM-like.  
	Consequently, the experimental searches in channels such as $h^0\to b\bar{b}$, $h^0\to \tau^+\tau^-$, $h^0\to Z^0Z^0$, and $h^0\to W^+W^-$ are insensitive to IDM parameters. 
	Only the loop-induced channels such as  $h^0\to \gamma \gamma$  \cite{ATLAS:2022tnm,CMS:2021kom} and $h^0\to \gamma Z^0$ \cite{ATLAS:2023yqk} 
	 can set constraints on the IDM scalar spectrum \cite{Arhrib:2012ia,Swiezewska:2012eh,Belanger:2015kga,Lundstrom:2008ai,Arhrib:2013ela,Abouabid:2023cdz}.  
	
	\item Furthermore, the invisible decay of the SM Higgs may be kinematically allowed if the dark matter candidate, either $H^0$ or $A^0$, is lighter than $m_{h^0}/2$. 
	This contributes to the total width of the SM Higgs boson. 
	ATLAS~\cite{ATLAS:2020kdi} and CMS~\cite{CMS:2022qva} have performed extensive searches for invisible Higgs decays.  
	The most recent bound, reported by ATLAS~\cite{ATLAS:2023tkt}, is:
	\begin{equation}
		\label{Ei18}
		\text{Br}_{\text{inv}} < 11\% \,\, \text{at} \,\, 95\% \,\, \text{C.L.}
	\end{equation}
	
	\item LEP searches also impose constraints on the masses of charged and neutral inert scalar bosons. 
	The corresponding limits are taken from Refs~\cite{Arhrib:2012ia,Swiezewska:2012eh,Belanger:2015kga,Lundstrom:2008ai}.  
	
	\item Finally, the IDM must be consistent with dark matter observations. 
	This includes relic density constraints, as well as limits from direct and indirect dark matter searches~\cite{Zyla:2020zbs, Abbott:1982af, Dine:1982ah, Preskill:1982cy,
		Belanger:2020gnr, Aprile:2018dbl, Amole:2019fdf, Abdelhameed:2019hmk,
		Agnes:2018ves, Aaboud:2017phn, Sirunyan:2017hci, Belyaev:2018ext, Lu:2019lok}.  
	
\end{itemize}

\section{Radiative corrections to: $\gamma \gamma \to H^{\pm}H^{\mp}$}
\label{sec:LO}
\subsection{Lowest order results}
%%%%%%%%%%%%%%%%%%%%%%%%%%%%%
%\subsection{Tree level results}

\begin{figure}[h]\centering
    \includegraphics[width=0.6\textwidth]{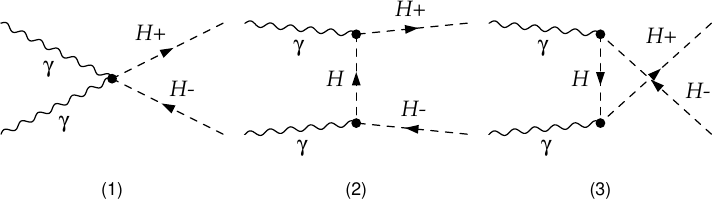}
    \caption{The tree-level Feynman diagrams for the process $\gamma\gamma \to H^\pm H^\mp$.}
    \label{fig:gghh}
\end{figure}

%%%%%%%%%%%%%%%%%%%%%%%%%%%%%
At the tree-level, the Feynman diagrams contributing to the process $\gamma\gamma \rightarrow H^\pm H^\mp$ are: 
the quartic $\gamma\gamma H^\pm H^\mp$ coupling diagram and the  t and u-channel diagrams as shown in Figure ~\ref{fig:gghh}.

Denoting the momentum of the incoming photons  and the outgoing charged scalar pair as $p_1$, $p_2$, $k_1$ and $k_2$, respectively.  
The Feynman rules needed to write down that amplitude are given in the Lagrangian Eq.~(\ref{cov-der}).

Meanwhile, in the center of mass energy frame, the momenta of incoming photons and the outgoing charged scalar bosons can be written as:
\begin{eqnarray}
    & & p_{1,2}=\frac{\sqrt{s}}{2} (1,0,0,\pm 1) \nonumber \\
    & & k_{1,2}=\frac{\sqrt{s}}{2} (1,\pm \beta_H \sin\theta,0,\pm \beta_H
    cos\theta) \nonumber
\end{eqnarray}
where $\sqrt{s}/2$ denotes the beam energy in a gamma-gamma  collider, and $\theta$ is the scattering angle
of  $H^-$ flying with reference to the laboratory frame, and $\beta_H = \sqrt{1-{\textstyle 4 m_{H^\pm}^2}/{\textstyle s}}$.

It is convenient to define the following Mandelstam variables:
\begin{eqnarray}
    & & s =  (p_1+p_2)^2 = (k_1+k_2)^2  \nonumber\\
    & & t = (p_1-k_1)^2 = (p_2-k_2)^2 = m_{H^\pm}^2- \frac{s}{2} (1   -\beta_H \cos\theta )  \nonumber\\
    & & u = (p_1-k_2)^2 = (p_2-k_1)^2 = m_{H^\pm}^2- \frac{s}{2} (1+\beta_H \cos\theta ) \nonumber \\
    & & s+t+u = 2 m_{H^\pm}^2  \nonumber
\end{eqnarray}

Using the above Feynman rules,  it is straightforward to write the lowest-order(LO) amplitude, which can have the following form:
\begin{equation}
    \begin{split}
        {\cal M}_1 =& 2 i e^2 g^{\mu \nu} \varepsilon_{\mu}\varepsilon_{\nu}\\
        {\cal M}_2 =& -i e^2\frac{(2k_1-p_1)_{\mu}(p_2-2 k_2)_{\nu}}{t-m_{H^{\pm}}^2}\varepsilon^{\mu}\varepsilon^{\nu}\\
        {\cal M}_3 =& -i e^2\frac{(p_1-2 k_2)_{\mu}(2 k_1-p_2)_{\nu}}{u-m_{H^{\pm}}^2}\varepsilon^{\mu}\varepsilon^{\nu}
    \end{split}
\end{equation}
%{\cal M}_2 =& -i e^2\frac{(2k_1-p_1)_{\mu}(k_1-p_1-k_2)_{\nu}}{t-m_{H^{\pm}}^2}\varepsilon^{\mu}\varepsilon^{\nu}\\
%{\cal M}_3 =& -i e^2\frac{(p_1-2 k_2)_{\mu}(p_1+k_1-k_2)_{\nu}}{u-m_{H^{\pm}}^2}\varepsilon^{\mu}\varepsilon^{\nu}

The squared amplitude, after summation over the polarizations of the incoming photons, gets the following form:
\begin{equation}
    \begin{split}
        \mid{\cal M}\mid^2=\mid{\cal M}_1 +{\cal M}_2+{\cal M}_3\mid^2=\frac{8e^4 \left(5 m_{H^{\pm}}^8-4 m_{H^{\pm}}^6 (t+u)+m_{H^{\pm}}^4 \left(t^2+u^2\right)+t^2 u^2\right)}{\left(m_{H^{\pm}}^2-t\right)^2 \left(m_{H^{\pm}}^2-u\right)^2}\\
        \frac{1}{4}\mid{\cal M}\mid^2=\frac{e^4}{4}(8 + \frac{(64 m_{H^{\pm}}^2 (4 m_{H^{\pm}}^2 + s (-1 + \cos^2\theta \beta_H^2)))}{ s^2 (-1 + \cos^2\theta \beta_H^2)^2})
    \end{split}
\end{equation}
The corresponding differential cross section  reads:
\begin{equation}
    \begin{split}
        \left(\frac{d \sigma}{d \cos\theta}\right)_{LO}=\frac{\alpha^2 \beta_H \pi}{8s}(8 + \frac{(64 m_{H^{\pm}}^2 (4 m_{H^{\pm}}^2 + s (-1 + \cos^2\theta \beta_H^2)))}{ s^2 (-1 + \cos^2\theta \beta_H^2)^2}).
    \end{split}
\end{equation}
Where $\alpha$ denotes the fine-structure constant. The total cross-section can be given as:
\begin{equation}
    \sigma_{LO}(\gamma\gamma \to H^\pm H^\mp)=\frac{2\alpha^2 \beta_H \pi}{s} (1 + \frac{16 m_{H^{\pm}}^4}{s^2 (1 - \beta_H^2)} + \frac{
        4 m_{H^{\pm}}^2 (s - 2 m_{H^{\pm}}^2) \log\frac{1 - \beta_H}{1 + \beta_H}}{ s^2 \beta_H})
    \label{eq:LO}
\end{equation}
The cross section Eq. (\ref{eq:LO}) of this process is only related to the mass of $H^{\pm}$ and is independent of the other four parameters $\mu^{2}_2, \lambda_{2}, m_{H^0}, m_{A^0}$ in the IDM.

\subsection{$\gamma\gamma \to H^\pm H^\mp$ at one-loop}
\label{sec:renormalization}

At tree level, the process $\gamma\gamma \to H^\pm H^\mp$ occurs through the scalar QED
interactions of the charged scalar boson:
(i) the contact vertex $\gamma \gamma H^\pm H^\mp$ and 
(ii) the $t$- and $u$-channel exchange of $H^\pm$ as shown in Figure ~\ref{fig:gghh}.
This defines the leading order cross section $\sigma_{\text{LO}}\equiv \sigma^{0}$, to which we add the complete NLO electroweak corrections.
We perform the one-loop calculation in the 't~Hooft--Feynman gauge, which simplifies the gauge structure while keeping all relevant cancellations manifest.
The complete set of generic one-loop topologies contributing to the amplitude is displayed in the Appendix (see Figs.~A.1--A.4). They include vertex corrections, box diagrams, self-energy insertions, counterterms and real photon emission:
\begin{itemize}
	\item Vertex corrections to the $\gamma H^+H^-$ and $Z^0H^+H^-$ couplings (Fig.~A.1),
	involving scalar and gauge boson loops. These diagrams renormalize the effective
	couplings to charged scalars.
	\item Box diagrams (Fig.~A.2), 
	%which provide finite and gauge-invariant fourth-order
	contributions to the scattering amplitude.
	\item Self-energies and counterterms (Fig.~A.3), which remove ultraviolet (UV) divergences
	in external legs and ensure a consistent renormalization of the electroweak parameters.
	\item Real-photon emission $\gamma\gamma \to H^\pm H^\mp \gamma$ (Fig.~A.4), required to cancel
	infrared (IR) divergences associated with soft virtual photons.
\end{itemize}

The renormalization pattern is simplified by the exact $\mathbb{Z}_2$ symmetry of the IDM,
which forbids $H^\pm$--$W^\pm$ and $H^\pm$--$G^\pm$ mixing.
%, since the $\gamma G^\pm H^\mp$ and $Z G^\pm H^\mp$ couplings are absent at tree level. 
The calculation is carried out using dimensional regularization~\cite{tHooft:1972tcz,Breitenlohner:1977hr}, with UV divergences removed in an on-shell renormalization scheme and IR singularities canceled by including soft and real-photon emission. 
This ensures that the final cross section is both UV- and IR-finite. 
A comprehensive treatment of IDM renormalization is given in Ref.~\cite{Banerjee:2019luv}. 
Here we adopt the on-shell framework originally developed for the SM~\cite{Bohm:1986rj,Hollik:1988ii,Denner:1991kt,Denner:2019vbn}, extended to the inert scalar sector so that both fields and their physical masses are consistently renormalized.

In the on-shell renormalization framework, inspired by established methodologies~\cite{Denner:1991kt}, we define the renormalization of the charged scalar field and its mass as follows:
\begin{eqnarray}
	H^{\pm} &\to& \sqrt{Z_{H^{\pm}}} H^{\pm} \approx \left(1 + \frac{1}{2} \delta Z_{H^{\pm}}\right) H^{\pm}, \nonumber\\
	m_{H^{\pm}}^2 &\to& m_{H^{\pm}}^2 + \delta m_{H^{\pm}}^2,
\end{eqnarray}
where $\delta Z_{H^{\pm}}$ is the wave function renormalization constant, and $\delta m_{H^{\pm}}^2$ is the mass counterterm. Additionally, we account for the renormalization of the photon wave function, the electric charge, and the electroweak mixing angle, following standard electroweak procedures~\cite{Denner:1991kt}.

Inserting  these definitions into the IDM Lagrangian, particularly the covariant derivative terms governing gauge-Higgs interactions, yields the following counterterms for the relevant vertices:
\begin{eqnarray}
	\delta \mathcal{L}_{\gamma H^{+} H^{-}} &=& -i e \left( \delta Z_e + \delta Z_{H^{\pm}} + \frac{1}{2} \delta Z_{\gamma\gamma} - \frac{c_W^2 - s_W^2}{4 c_W s_W} \delta Z_{Z^0\gamma} \right) (k_1 - k_2)_{\mu} A^{\mu} H^{+} H^{-}, \nonumber \\
	\delta \mathcal{L}_{\gamma \gamma H^{+} H^{-}} &=& 2 i e^2 \left( 2 \delta Z_e + \delta Z_{H^{\pm}} + \delta Z_{\gamma\gamma} + \frac{s_W^2 - c_W^2}{2 c_W s_W} \delta Z_{Z^0\gamma} \right) g_{\mu \nu} A^{\mu} A^{\nu} H^{+} H^{-},
\end{eqnarray}
where $k_1$ and $k_2$ are the momenta of the outgoing $H^+$ and $H^-$, respectively, $\delta Z_e$ is the electric charge renormalization constant, $\delta Z_{\gamma \gamma}$ is the photon wave function renormalization, and $\delta Z_{Z^0\gamma}$ accounts for photon-$Z^0$ mixing. These counterterms ensure the cancellation of UV divergences in the one-loop amplitudes.

The on-shell conditions fix the renormalization constants by imposing physical constraints on the charged scalar propagator:
\begin{eqnarray}
	&&\text{Re} \left. \frac{\partial \hat{\Sigma}^{H^{\pm} H^{\mp}}(k^2)}{\partial k^2} \right|_{k^2 = m_{H^{\pm}}^2} = 0, \nonumber \\
	&&\text{Re} \hat{\Sigma}^{H^{\pm} H^{\mp}}(m_{H^{\pm}}^2) = 0,
\end{eqnarray}
where $\hat{\Sigma}^{H^{\pm} H^{\mp}}$ is the renormalized self-energy of the charged scalar  boson. These conditions ensure that the pole mass remains at $m_{H^{\pm}}$ and that the residue of the propagator is unity, consistent with a physical particle interpretation.

For the gauge sector, the renormalization constants for the photon, $Z^0$ boson, their mixing, and the Weinberg angle are determined following electroweak on-shell prescriptions~\cite{Denner:1991kt}. 

%{\color{red}
For the renormalization of charge, it is firstly done under zero-momentum-transfer scheme (also called on-shell scheme in some literature) in the Thomson Limit,
and then converted into $\alpha(m_{Z^0})$ scheme which absorbs large logarithmic contributions from light fermions into the running coupling constant, improving perturbative convergence~\cite{Denner:1991kt,Sun:2016bel,Xie:2018yiv}. 
The electric charge renormalization constant in this scheme is related to the zero-momentum-transfer scheme via:
%in which the large logarithm from fermions are also absorbed into the redefinition of running coupling constant~\cite{Denner:1991kt,Sun:2016bel,Xie:2018yiv}.
%The corresponding renormalization constant is related to zero-momentum-transfer scheme via:
\begin{equation}
    \delta Z_e|_{\alpha(m_{Z^0})}=\delta Z_e|_{\alpha(0)}-\dfrac{1}{2}\Delta\alpha(m_{Z^0})
\end{equation}
with
\begin{equation}
    \Delta\alpha(m_{Z^0})=\Pi^{\gamma \gamma}_{f\neq \mathrm{top}}(0)-\mathrm{Re}\Pi^{\gamma \gamma}_{f\neq \mathrm{top}}(m_{Z^0}^2)
\end{equation}
and
\begin{equation}
    \Pi^{\gamma \gamma}(s)=\frac{\partial\hat{\Sigma}^{\gamma \gamma}}{\partial k^{2}}\biggr|_{k^2=s}
 %   -\mathrm{Re}\Pi^{AA}_{f\neq \mathrm{top}}(m_Z^2)
\end{equation}
Here: $\hat{\Sigma}^{\gamma \gamma}$ represents the renormalized photon self-energy, $\alpha(0)$ denotes the running coupling constant in the zero-momentum transfer scheme, and $f\neq \mathrm{top}$ denotes only the fermionic (without top quark) contribution is included in the self-energy of the photon.
The running coupling constant is then defined as
\begin{equation}
    \alpha(m_{Z^0})=\dfrac{\alpha(0)}{1-\Delta\alpha(m_{Z^0})}.
\end{equation}

%In this analysis, we adopt the $\alpha(m_Z)$ scheme, which absorbs large logarithmic contributions from light fermions into the running coupling constant, improving perturbative convergence~\cite{Denner:1991kt,Sun:2016bel,Xie:2018yiv}. The electric charge renormalization constant in this scheme is related to the $\alpha(0)$ scheme via:
%\begin{equation}
%	\delta Z_e^{\alpha(m_Z)} = \delta Z_e^{\alpha(0)} - \frac{1}{2} \Delta \alpha(m_Z),
%\end{equation}
%where the photon vacuum polarization  function is given by:
%\begin{equation}
%	\Delta \alpha(m_Z) = \Pi^{AA}_{f \neq \text{top}}(0) - \text{Re} \Pi^{AA}_{f \neq \text{top}}(m_Z^2).
%\end{equation}
%The running coupling is then defined as:
%\begin{equation}
%	\alpha(m_Z) = \frac{\alpha(0)}{1 - \Delta \alpha(m_Z)}.
%\end{equation}
This choice ensures that the electroweak corrections are well-controlled, particularly for processes sensitive to the IDM’s charged scalar sector at high-energy scales relevant to photon colliders.

We now turn to the issue of IR divergences. These arise from two distinct sources:
virtual photon loop corrections and the emission of real photons. Virtual photon exchange generates soft infrared divergences. We regulate them by introducing
a small fictitious photon mass $\lambda$ in the loop integrals. In parallel, the real emission
process $\gamma\gamma\to H^\pm H^\mp \gamma$ is split into a soft region ($E_\gamma<\Delta E$) and
a hard region ($E_\gamma\geq\Delta E$), with $\Delta E=\delta_s\sqrt{s}/{2}$ an arbitrary energy cutoff in the
$\gamma\gamma$ center-of-mass frame. The soft part factorizes universally over the Born cross
section and contains $\ln(2\Delta E/\lambda)$ terms, which cancel exactly against the
$\ln\lambda$ terms from the virtual corrections, in accordance with the Kinoshita--Lee--Nauenberg (KLN) theorem \cite{Kinoshita:1962ur,Lee:1964is}. The remaining $\Delta E$
dependence cancels once the hard real emission contribution is included.
Note that we have verified, with high numerical accuracy, the cancellation of the fictitious photon mass $\lambda$. In addition, we explicitly tested the stability of our predictions under variations of the soft-photon cutoff $\Delta E$, and, as shown in Appendix \ref{check:de}, the cross section remains independent of this unphysical parameter. 
%\footnote{No angular cutoff $\Delta\theta$ is needed for our process. Initial-state collinear singularities are absent because photons carry no charge, while final-state collinear singularities are regulated by the charged scalar mass $m_{H^\pm} \neq 0$.}.\\

%{\color{red}
On the other hand, in order to deal with the large logarithm terms from light fermions, the structure function approach~\cite{Kuraev:1985hb} is applied again as in our former work~\cite{Abouabid:2022rnd}. 
%Finally, the NLO corrections are assembled as:
%At next-to-leading order, the corrections naturally split into three parts: the
%virtual loop contribution, soft real emission, and hard real emission. 
%The NLO cross section can therefore be written as
Finally, the NLO cross section can be written as follows
\begin{equation}
	d\sigma^{1}
	= d\sigma_{V}(\lambda) + d\sigma_{S}(\lambda,\Delta E) + d\sigma_{H}(\Delta E) + d\sigma_{CT}\,,
	\label{fullcor}
\end{equation}
where:
\begin{itemize}
	\item $d\sigma_{V}(\lambda)$ denotes the virtual one-loop contribution, which is
	infrared-divergent and regulated by introducing a fictitious photon mass $\lambda$.
	\item $d\sigma_{S}(\lambda,\Delta E)$ corresponds to soft-photon emission with
	$E_\gamma < \Delta E$, and contains $\ln\lambda$ terms that cancel against those
	in the virtual corrections.
	\item $d\sigma_{H}(\Delta E)$ represents the hard-photon emission with
	$E_\gamma \geq \Delta E$, which is finite.
	\item $d\sigma_{CT}$ denotes the ``counter term'' from structure function approach, which is given by
\begin{equation}
%    \begin{aligned}
        d \sigma_{CT}= \frac{2\alpha}{3 \pi}d\sigma^0   \sum_{f} N_c^f Q_f^2 \log \frac{s}{4 m_{f}^{2}} .
%    \end{aligned}
\end{equation}
\end{itemize}
The differential cross section for soft photon emission ($E_\gamma < \Delta E$) is integrated analytically:
\begin{equation}
    \begin{aligned}
        d\sigma_S= & -\frac{\alpha}{\pi}d\sigma^0 \times\biggl\{2 \log \frac{2 \Delta E}{\lambda}+\frac{1+\beta_H^{2}}{\beta_H} \log \frac{2 \Delta E}{\lambda} \log \left(\frac{1-\beta_H}{1+\beta_H}\right)+\frac{1}{\beta_H} \log \left(\frac{1-\beta_H}{1+\beta_H}\right)
        \\& +\frac{1+\beta_H^{2}}{\beta_H}\left[\operatorname{Li}_{2}\left(\frac{2 \beta_H}{1+\beta_H}\right)+\frac{1}{4}\log ^{2}\left(\frac{1-\beta_H}{1+\beta_H}\right)\right]\biggr\} .
    \end{aligned}
\end{equation}
%{\color{blue}
	%\subsubsection{structure function}
%	\begin{equation}
%		\begin{aligned}
%			d\sigma_S(\gamma\gamma\rightarrow H^\pm H^\mp )= & -\frac{\alpha}{\pi}d\sigma_0(\gamma\gamma\rightarrow H^{\pm}H^{\mp})
%			\\&\times\{2 \log \frac{2 \Delta E}{\lambda}+\frac{1+\beta_H^{2}}{\beta_H} \log \frac{2 \Delta E}{\lambda} \log \left(\frac{1-\beta_H}{1+\beta_H}\right)
%			\\ &+\frac{1}{\beta_H} \log \left(\frac{1-\beta_H}{1+\beta_H}\right)
%			\\ &+\frac{1+\beta_H^{2}}{\beta_H}\left[\operatorname{Li}_{2}\left(\frac{2 \beta_H}{1+\beta_H}\right)+\frac{1}{4}\log ^{2}\left(\frac{1-\beta_H}{1+\beta_H}\right)\right]\}
%		\end{aligned}
%	\end{equation}

	The hard photon contribution ($E_\gamma > \Delta E$) is infrared finite and evaluated
	numerically from the factorized phase space:
	\begin{equation}
		\sigma_{\rm hard}(\Delta E) =
		\frac{1}{2s} \int_{E_\gamma > \Delta E}
		\frac{d^3 \mathbf{k}_\gamma}{(2\pi)^3 2E_\gamma}\;
		\overline{|\mathcal{M}\bigl(\gamma\gamma \to H^\pm H^\mp \gamma \bigr)|^2}\;
		d\Phi_{H^+H^-}\, ,
		\label{eq:sigmaHard}
	\end{equation}
	where $d\Phi_{H^+H^-}$ is the two-body phase space of the charged scalar pair recoiling
	against the emitted photon. The bar indicates averaging over the initial photon
	polarizations and summing over the final-state photon polarizations.
	
	It is worth noting that near threshold, the long-range electromagnetic interaction between $H^+$ and $H^-$ induces the well-known Coulomb enhancement, appearing at one-loop as a singular correction proportional to $1/\beta_H$. 
	Since such terms grow large close to threshold, they must be resummed to all orders using the Sommerfeld factor. 
	This procedure yields a finite and reliable prediction, properly accounting for multiple soft-photon exchanges between the charged scalars. 
	For further details on the explicit prescription adopted for this singularity, we refer the reader to our previously published work \cite{Abouabid:2022rnd}.

The total NLO cross section, $\sigma^{\text{NLO}}$, can be written as the
sum of the leading-order contribution, $\sigma^{0}$, and the one-loop correction, $\sigma^{1}$,
\begin{eqnarray}
	\sigma^{\text{NLO}} &=& \sigma^{0} + \sigma^{1} 
	\;\equiv\; \sigma^{0}\,(1+\Delta)\,,
	\label{sig1}
\end{eqnarray}
where $\Delta$ denotes the relative correction factor. 
%It is convenient to decompose $\Delta$
%into two separately gauge-invariant pieces,
%\begin{eqnarray}
%	\Delta \;=\; \Delta_{\text{weak}} + \Delta_{\text{QED}}\,,
%	\label{split}
%\end{eqnarray}
%corresponding to the purely weak and the QED contributions, respectively.

%{\color{red}
As described in Section 3.1 of Ref.\cite{Beenakker:1991ca}, the NLO electroweak corrections can be safely grouped into two gauge-invariant parts:
\begin{enumerate}
\item The ``QED'' part, which includes all the diagrams that contain an extra photon attached to the LO diagrams and the photonics contribution to the wave-function renormalization of the charged scalar.
\item The ``weak'' part, which contains all the remaining contributions.
\end{enumerate}
It should be noted that the ``QED'' part here is only a gauge-invariant subgroup of the whole QED corrections since it is very hard to separate the whole QED part due to $\gamma-Z^0$ mixing.
The relative correction is then separated into two parts correspondingly, i.e. we have
\begin{equation}
\Delta =\Delta_{\mathrm{weak}}+\Delta_{\mathrm{QED}}.
\label{split}
\end{equation}
%}

For the evaluation of one-loop amplitudes and counterterms we use the
\texttt{FeynArts} and \texttt{FormCalc} packages~\cite{Hahn:2000kx,Hahn:1998yk,Hahn:2006qw}, while scalar
integrals are computed with \texttt{LoopTools}~\cite{Hahn:1999mt,Hahn:2010zi}.
We have explicitly verified UV-finiteness both analytically and numerically. 
Independent cross-checks with the \texttt{FDC} framework~\cite{Wang:2004du} further confirm the correctness of our results.

\section{Numerical results}
\label{sec:results}
In this section, we present our numerical findings for the process $\gamma \gamma \to H^\pm H^\mp$ taking into account full radiative corrections. 

To determine our numerical results, we have adopted the following physical parameter values from PDG\cite{Tanabashi:2018oca} as input:
\begin{enumerate}
    \item The fine structure constants at renormalization scales $\mu=0$ and $\mu=m_{Z^0}$ are taken as $\alpha(0)=1/{137.036}$ and $\alpha(m_{Z^0})={1}/{128.943},$ where the 5 flavor quark scheme is used with $ \Delta \alpha^{(5)}_{hadron}(m_{Z^0})=0.02764 $.
    \item The weak gauge boson masses are used with $m_{W}=80.38$ GeV and $m_{Z^0}=91.19$ GeV.
    \item The charged lepton masses are taken as $m_{e}=0.511 \text{ MeV, } $ $m_{\mu}=0.105 \text{ GeV, }$  $m_{\tau}=1.77 \text{ GeV.}$ The quark masses are taken as $m_u=m_d=3.45\text{ MeV, }$ $m_c=1.275\text{ GeV, }$  $m_s=0.095\text{ GeV, } $  $m_{t}=173.0 \text{ GeV, }$ $ m_b=4.66 \text{ GeV, } $
\end{enumerate}
We have taken $m_{h^0}=125.18$ GeV in the IDM and have assumed that the CP even Higgs boson $h^0$ is the discovered Higgs like particle by the LHC collaborations. We have conducted a systematic scan for the free IDM parameters, which include the physical masses $m_{H^0}$, $m_{A^0}$, and $m_{H^\pm}$, as well as the parameters $\lambda_2$ and $\mu_{2}^2$. We have taken into account the LHC and LEP limits for new scalar particles as well as all theoretical constraints listed in the subsection \ref{sec:thexcons}.

It is found that our numerical results are almost independent of $\lambda_2$. Thus, for the sake of simplicity, we fix  $\lambda_2=2$ in what follow.

For the convenience of our numerical analysis, we define three scenarios in Table \ref{tab:scenarios}. These scenarios have different theoretical hypothesis and different experimental constraints. Scenario I adopts the degenerate spectrum in the second  doublet and is the simplest one with only three free parameters and is the easiest one to demonstrate the one-loop corrections. While Scenario III might be more realistic after taking into account collider experimental bounds and  dark matter constraints and seven benchmark points are chosen from this scenario for the LHC and future collider searches. In the Scenario II, the experimental constraints of dark matter have not been taken into account, when compared with the Scenario III.

\begin{table}[!htb]
	\centering
	\begin{tabular}{|c|c|c|c|c|c|}
		\hline
		& Scenario I & Scenario II & Scenario III  \\
		\hline
		Theoretical constraints & $\checkmark$ & $\checkmark$& $\checkmark$\\
		\hline 
		Degenerate spectrum &$\checkmark$&&\\
		\hline
		\hline
		Higgs data &$\checkmark$&$\checkmark$&$\checkmark$\\
		\hline
		Higgs invisible decay open &&$\checkmark$&$\checkmark$\\
		\hline
		Direct searches from LEP &$\checkmark$&$\checkmark$&$\checkmark$ \\
		\hline
		Electroweak precision tests & $\checkmark$&$\checkmark$&$\checkmark$\\
		\hline
		Dark matter constraints & & &$\checkmark$\\
		\hline
	\end{tabular}
	\caption{Three scenarios are defined and their conditions are tabulated. Scenario I assuming all the inert scalars are degenerate is described by only three parameters which are $(m_S=m_{H^0}=m_{A^0}=m_{H^\pm},\mu_2^2,\lambda_{2})$, scenarios II and III possessing a non-degenerate spectrum for the inert scalars are described by five parameters, which are  $(m_{H^\pm}, m_{H^0}, m_{A^0},\mu_2^2, \lambda_2)$.}
	\label{tab:scenarios}
\end{table}

%we choose $\Delta \theta=\Delta E=10^{-3}$ as our choose.

%\subsection{$\gamma \gamma \rightarrow H^+H^-$}

%In this section we discuss radiative corrections to charged Higgs pairs. 
We first present the LO/NLO cross sections of $\gamma \gamma \to H^\pm H^\mp$ as a function of the charged scalar mass in the scenario I for 250 GeV, 500 GeV and 1 TeV CM energy,  respectively. Five cases are chosen to demonstrate the effects of theoretical and experimental constraints, the size of weak corrections, and the size of real QED corrections as well. The parameter  $\mu_2^2$ is given in Table \ref{tab:idms}.

\begin{table}[!htb]
    \centering
    \begin{tabular}{|c|c|c|c|c|c|}
        \hline
                           & IDM1  & IDM2 & IDM3 & IDM4   & IDM5   \\
        \hline
        $\mu_2^2$(GeV$^2$) & 40000 & 6000 & 0    & -10000 & -30000 \\
        \hline
        %		$\mathrm{Lower}$(GeV$^2$)  & $h^0\rightarrow \gamma\gamma$ & N.A.  & N.A.  & $h^0\rightarrow \gamma\gamma$ & $h^0\rightarrow \gamma\gamma$ \\
        %		\hline
        %		$\mathrm{Upper}$(GeV$^2$)  & N.A. & N.A.  & N.A.  & N.A & Unitarity \\
        %		\hline
    \end{tabular}
    \caption{In Scenario I, five cases with typical values of $\mu_2^2$  labelled as IDM1-5 are given and will be used in Fig.  \ref{ggHH-degen}.}
    \label{tab:idms}
\end{table}

\begin{figure}[H]
    \begin{center}
        \includegraphics[width=0.3\textwidth]{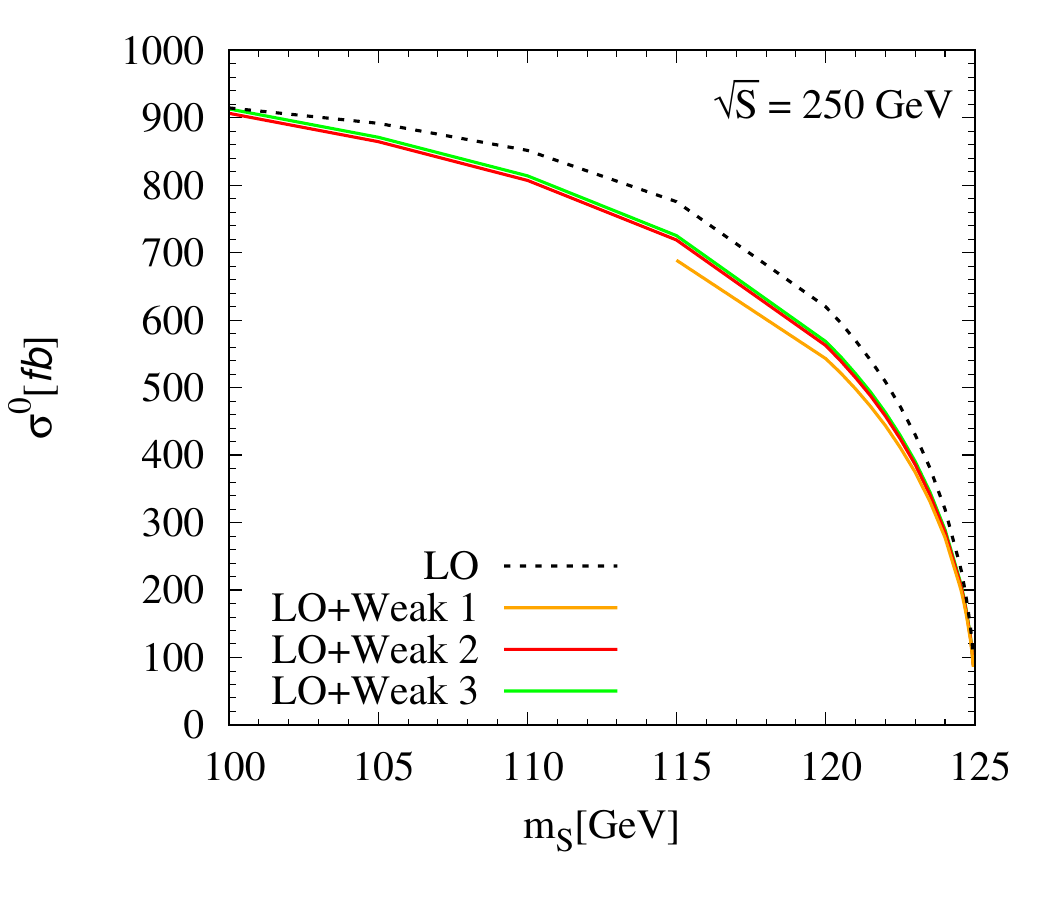}
        \includegraphics[width=0.3\textwidth]{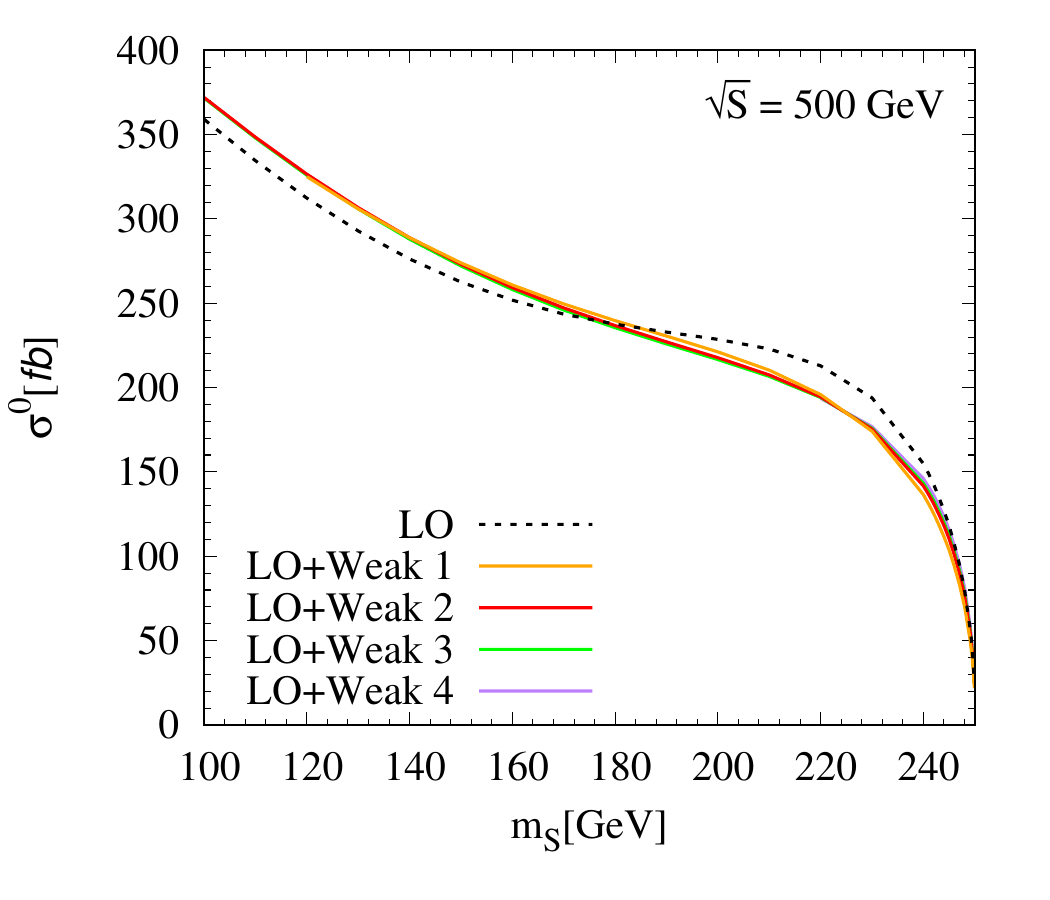}
        \includegraphics[width=0.3\textwidth]{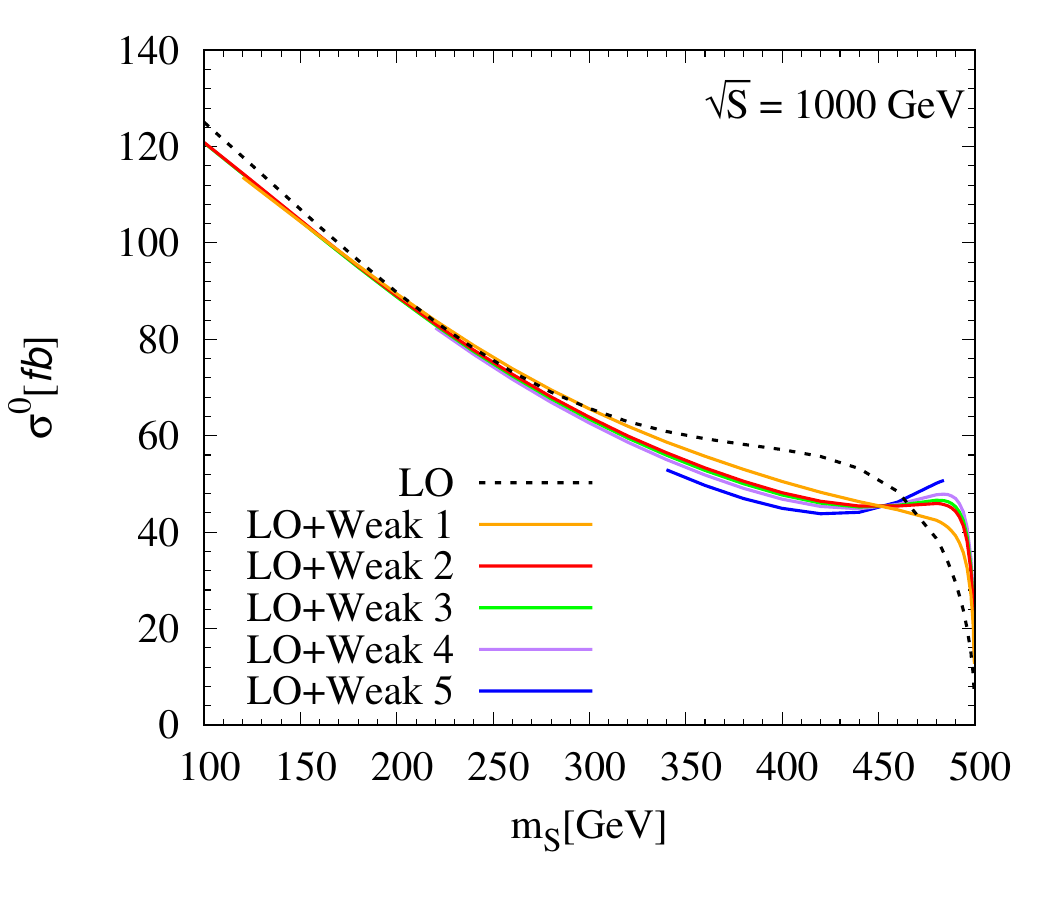}\\
        \includegraphics[width=0.3\textwidth]{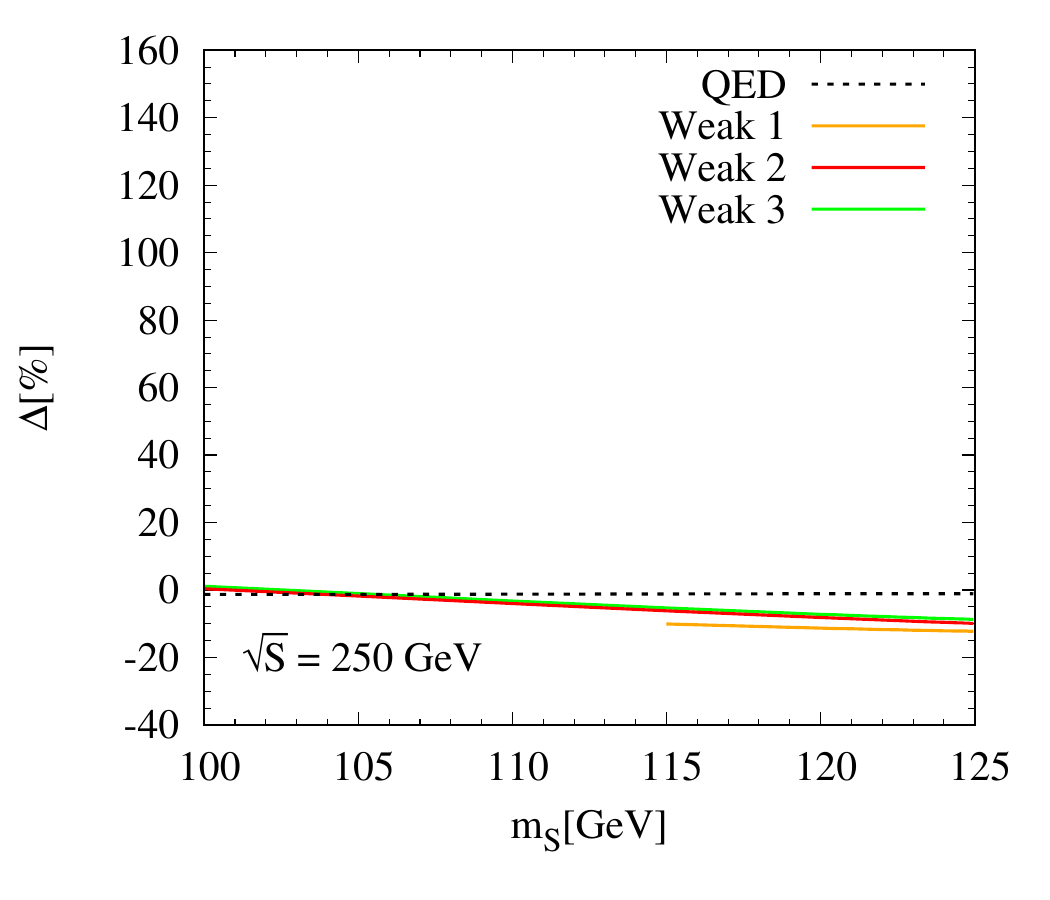}
        \includegraphics[width=0.3\textwidth]{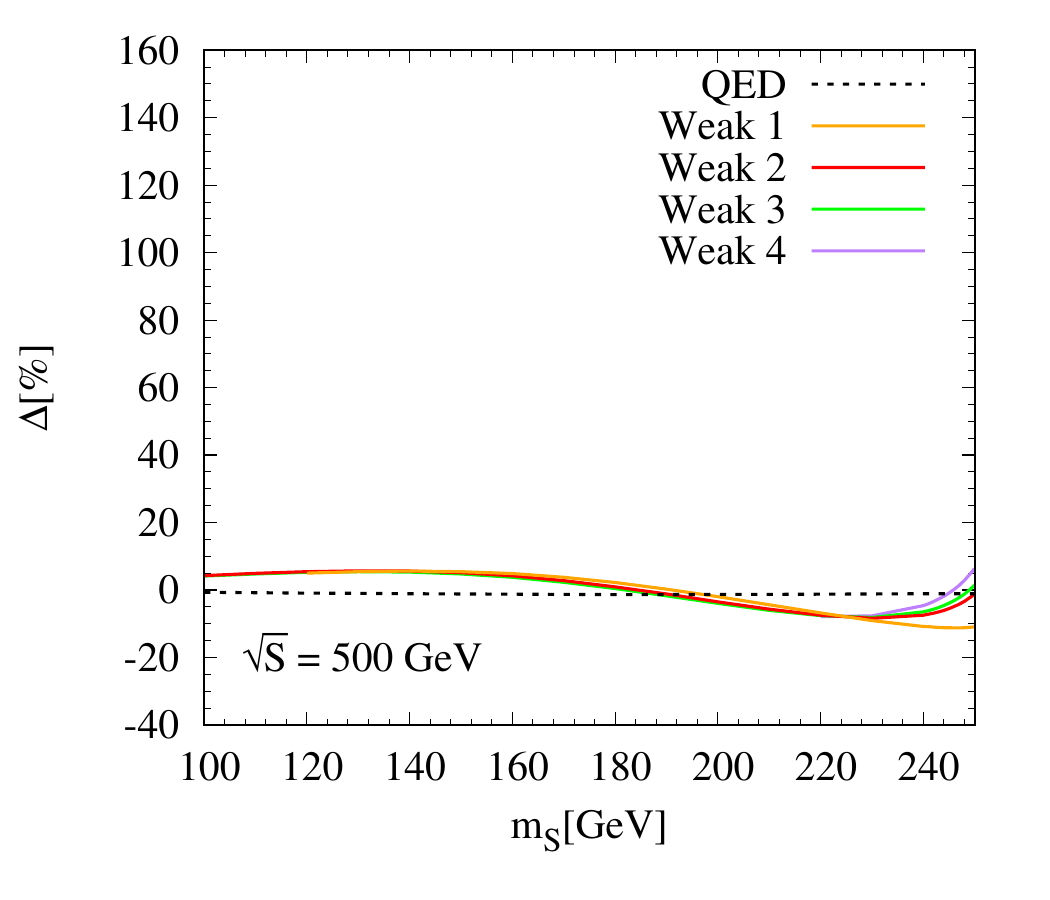}
        \includegraphics[width=0.3\textwidth]{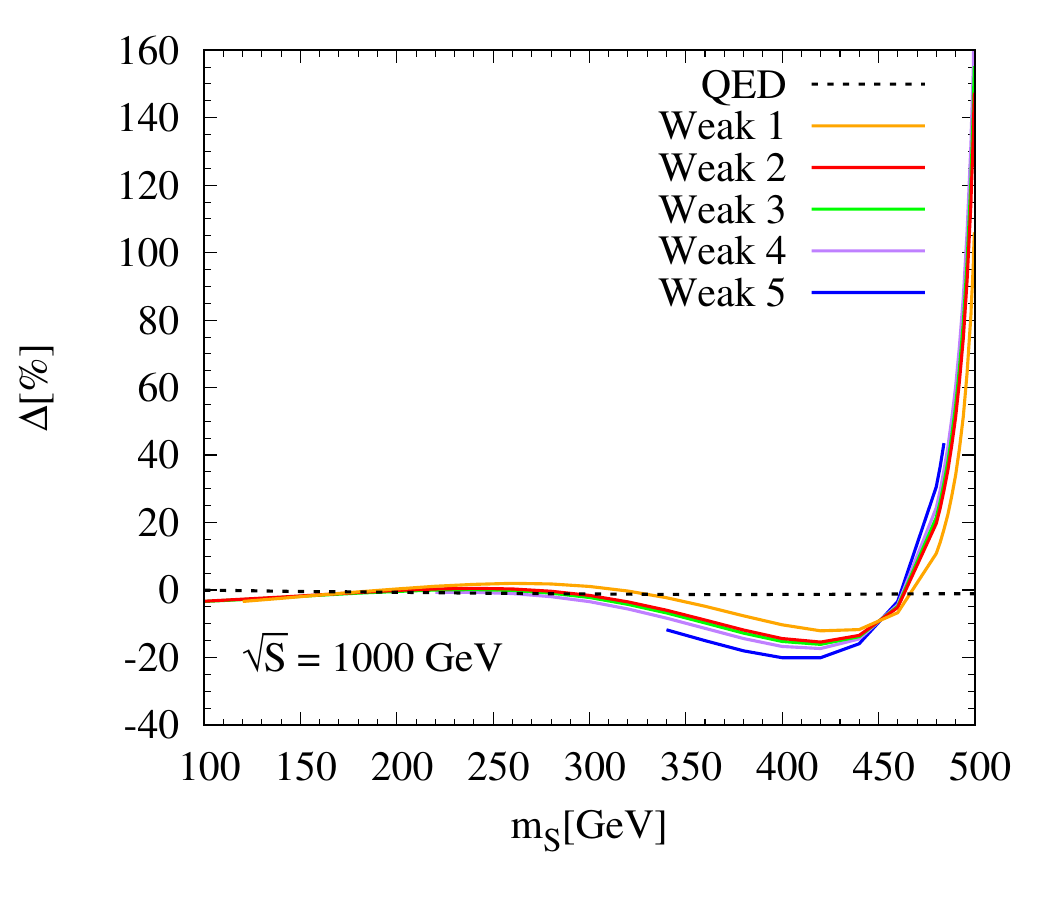}
    \end{center}
    \caption{Total cross sections and the percentage of corrections for the process $\gamma\gamma \to H^{\pm} H^{\mp}$ with three collision energies $\sqrt{s}=250$ GeV,  $500$ GeV, and $1000$ GeV are demonstrated, respectively. As an example, the LO result and the full NLO results of IDM in the scenario I are shown. The full NLO results include the LO, one-loop weak corrections, and the real emission corrections. The x-axis denotes the mass of charged scalar boson mass, and the y-axis denotes the total cross section (upper panel) and the percentage of corrections $\Delta$ (lower panel) given in Eq. (\ref{split}).}
    \label{ggHH-degen}
\end{figure}

The numerical results are presented in Fig. \ref{ggHH-degen}, where both the cross sections and percentages of NLO corrections, given in Eq. (\ref{split}), are shown.

The cross sections and the percentages of NLO corrections for those five case introduced in Table \ref{tab:idms} are shown in Fig. \ref{ggHH-degen}. From the plots, it is easy to read out that for the collision energy $\sqrt{s}=250$ GeV, only the first three cases given in Table \ref{tab:idms} could be probed; while for $\sqrt{s}=500$ GeV, the first four cases can be probed. Only when $\sqrt{s}=1$ TeV, all these five cases can be probed. 

These plots demonstrate that the cross sections depend upon the mass of charged scalar boson and other model parameters like $\mu_2^2$, while the five cases given in Table \ref{tab:idms} are reflected in the ranges of the mass of charged scalar boson. The NLO corrections can be either positive and negative and the size can reach  $10\%$ or so, as clearly shown by the plots at   $\sqrt{s}=500$ GeV.

There are two more points which are noteworthy.
\begin{itemize}
\item The cross section of $\gamma\gamma \to H^{\pm} H^{\mp}$ with $m_{H^\pm}=100$ GeV at $\sqrt{s}=250$ GeV is typically 900 fb. With the increase of collision energy reach to $1$TeV, the cross section drops down to 120 fb. It does not drop with a law like $1/s$. Such a feature can be attributed to the fact that the leading process of $\gamma\gamma \to H^\pm H^\mp$ includes both contact interaction and t/u-channels.
\item The weak corrections can be either positive or negative for $\gamma\gamma \to H^\pm H^\mp$, which is typically dependent upon the collision energy and the mass of charged scalar boson. A salient feature is that  the magnitude can vary from a few percent to even $+100\%$ (at $\sqrt{s}=1$ TeV when the mass of charged scalar boson is close to the kinematic threshold, when $m_{H^\pm}=495$ GeV. 
\end{itemize}

\begin{figure}[H]\centering
    \includegraphics[width=0.31\textwidth]{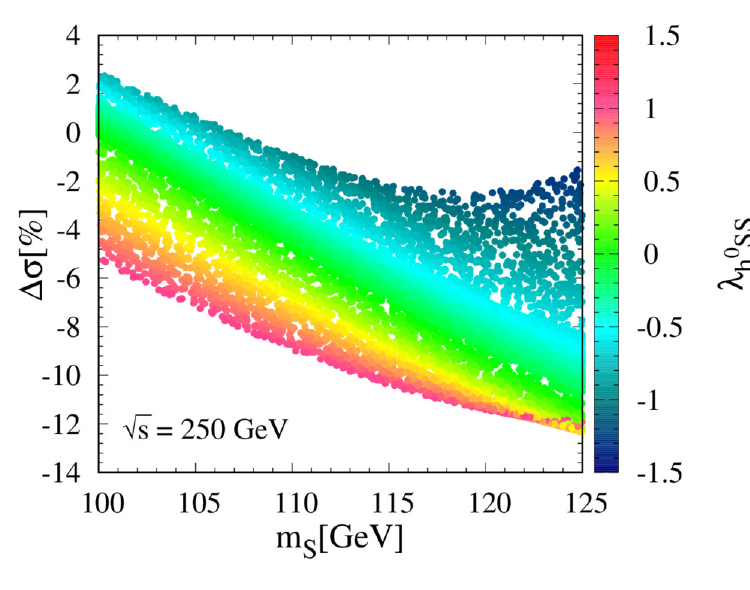}
    \includegraphics[width=0.31\textwidth]{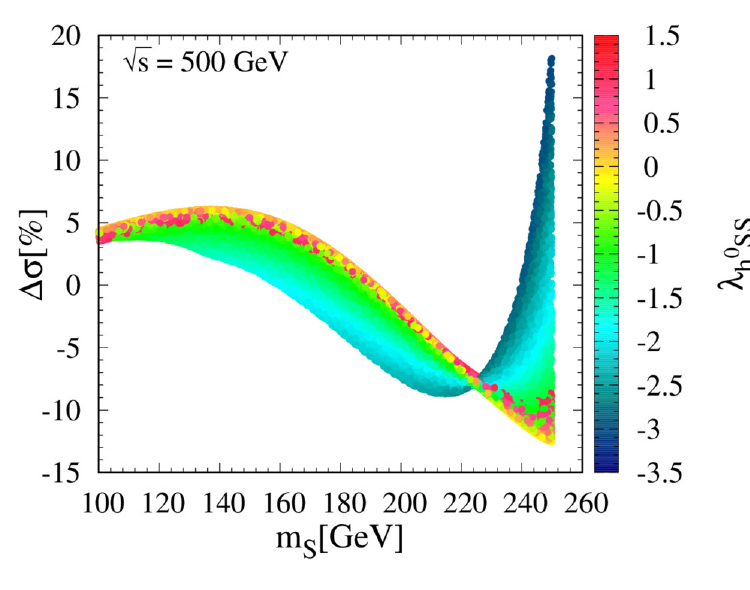}
    \includegraphics[width=0.31\textwidth]{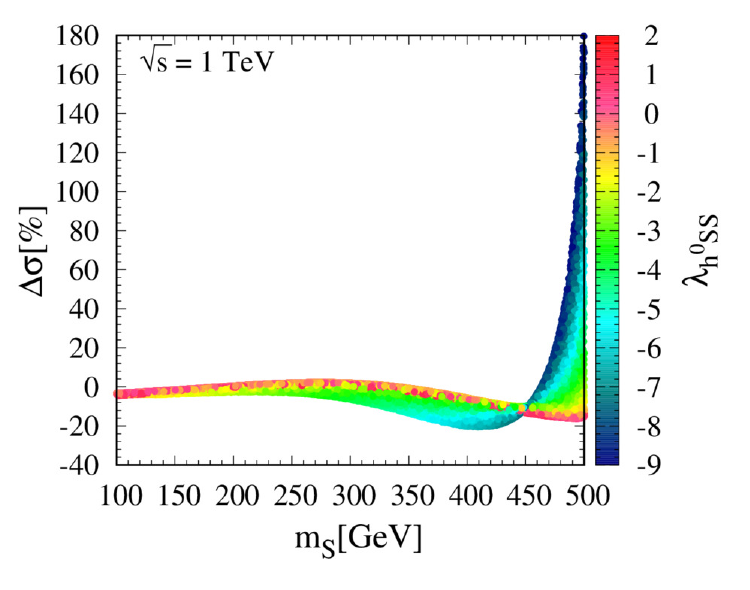}\\
    \includegraphics[width=0.31\textwidth]{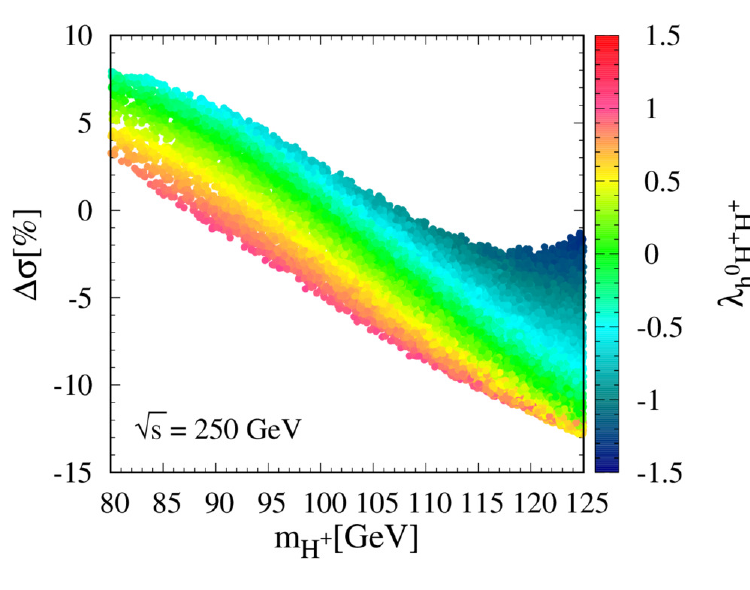}
    \includegraphics[width=0.31\textwidth]{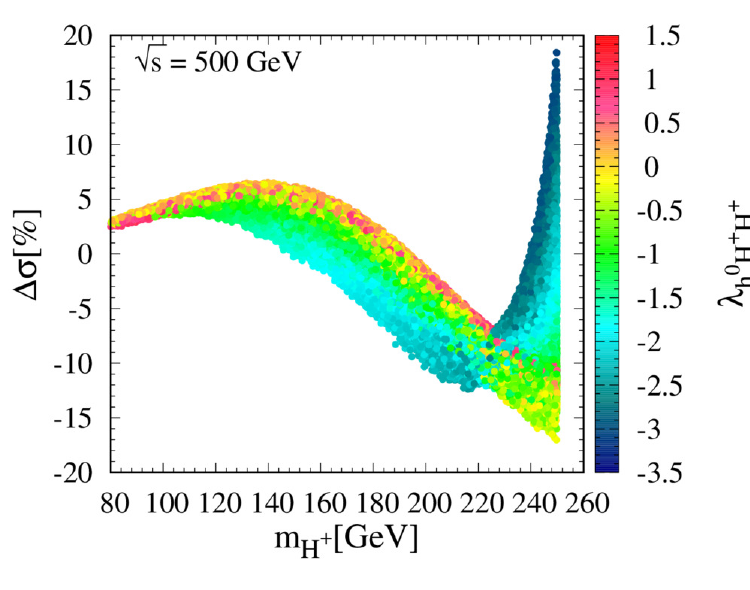}
    \includegraphics[width=0.31\textwidth]{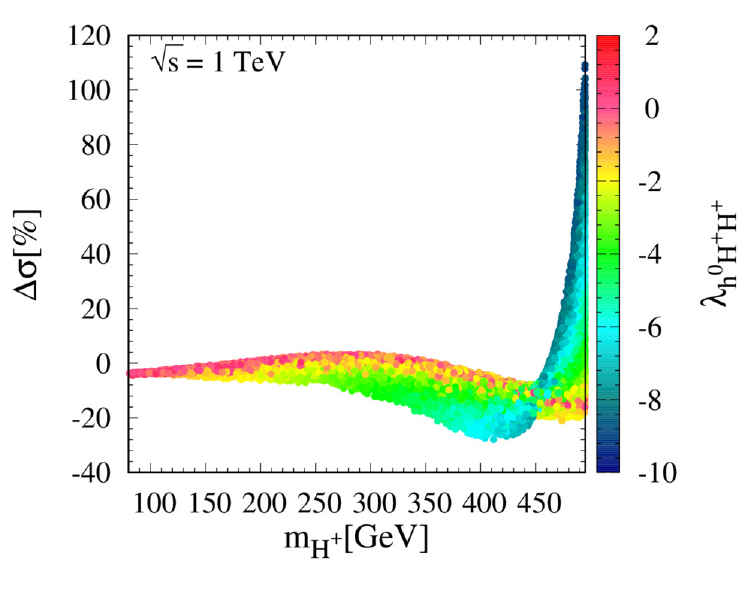}\\
    \includegraphics[width=0.31\textwidth]{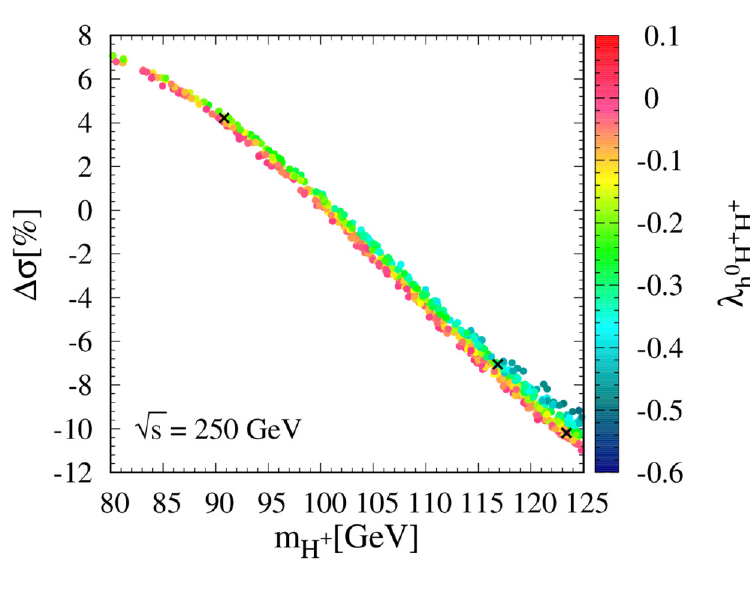}
    \includegraphics[width=0.31\textwidth]{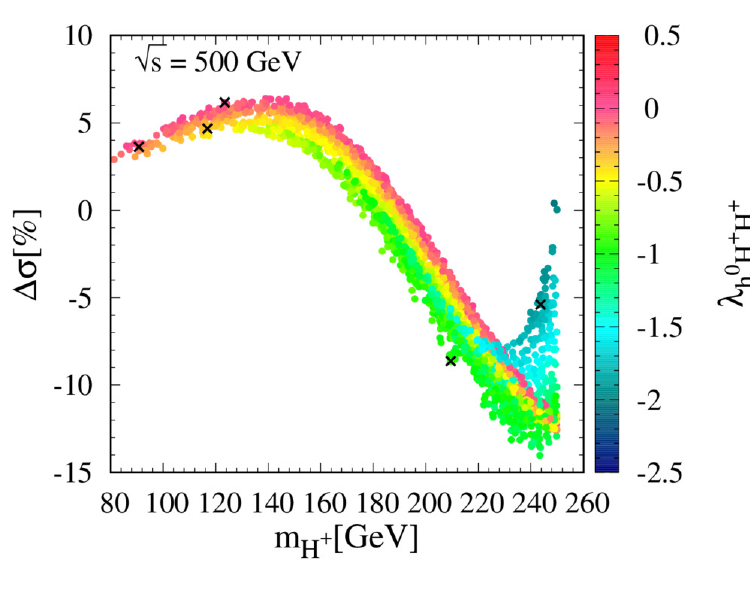}
    \includegraphics[width=0.31\textwidth]{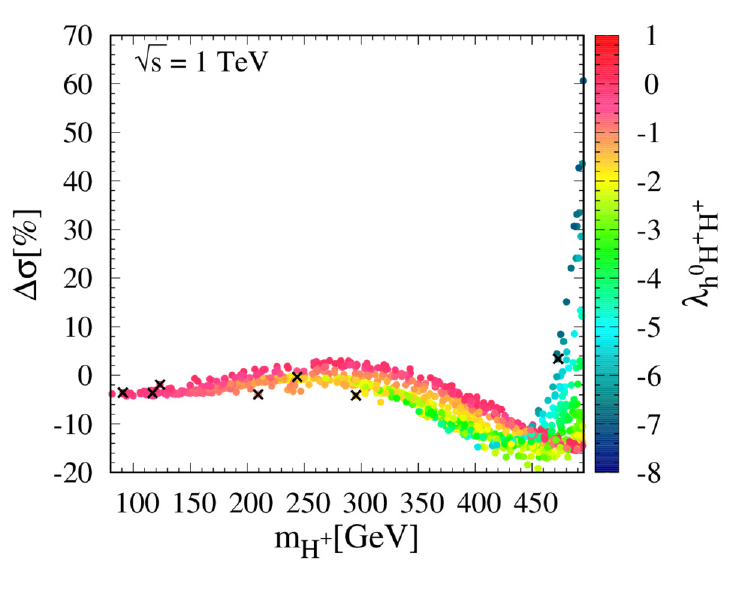}
    \caption{Electroweak corrections to  $\gamma \gamma \to H^\pm H^\mp$ for $\sqrt{s}=250$ GeV, 500 GeV and 1000 GeV  as a function of $m_S$  and the triple Higgs couplings $\lambda_{h^0SS}$ normalized to the SM Higgs vev. Upper panels show the degenerate scenario, middle and lower panels are respectively for the non-degenerate scenario before and after applying dark matter constraint. }
    \label{fig:ggHH}
\end{figure}

In Fig. \ref{fig:ggHH}, the percentage of corrections given in Eq. (\ref{split}) for three scenarios are presented. 
In upper panel of the Fig.\ref{fig:ggHH}, we depict the relative corrections in the degenerate mass scenario to $\gamma \gamma \to H^\pm H^\mp$ as a function of $m_S=m_{H^0}=m_{A^0}=m_{H^\pm}$ where the color palette represents the triple couplings $\lambda_{h^0SS}/v=-\lambda_3$ scaled by the vacuum expectation value. 
Similar conventions are applied to the results of the other two scenarios, which are shown in the middle and lower panel, respectively. 
When the results of scenario II and scenario III are compared, it is noteworthy that the dark matter constraint can kill more than 95$\%$ points of the parameter space. 

A remarkable feature is that near the threshold region for the collision energy $\sqrt{s}=1$ TeV, the corrections can reach to $+180\%$ for scenario I, $+120\%$ for scenario II, and $+60\%$ for scenario III. These large corrections can occur at the parameter region where the absolute value of trilinear Higgs coupling $|\lambda_{h^0 H^+ H^-}|$ is large, i.e. when the charged scalar boson is heavy (near 500 GeV). For the large value of $\lambda_{h^0 H^+ H^-}$ coupling, the significant enhancement on the radiative corrections could be attributed to the box diagrams with charged scalar exchange which would have $\lambda_{h^0 H^+ H^-}^2$ dependency.

In contrast, when the collision energy is $\sqrt{s}=250/500$ GeV, the corrections can not be as sizable as in the case of $\sqrt{s}=1$ TeV, the underlying is that when collision energy is lower than $1$ TeV, the colliders is not able to probe the parameter region with a large absolute value of trilinear Higgs coupling $|\lambda_{h^0 H^+ H^-}|$, which is determined by $\lambda_3$, or equivalently by the mass of charged scalar boson mass as given in Eq. (\ref{eq:lams}). But the size of corrections can reach $\pm10\% \sim \pm 20\%$ in the allowed parameter space.

In the scenario III, the black cross markers denote the benchmark points given in Table \ref{tab:BPs}, which are consistent with collider experiments and dark matter searches as proposed in Ref \cite{Abouabid:2022rnd}. Here in this work, only BP1 is a new benchmark point. Mass spectra and major decay tables are provided. The related phenomenologies for benchmark points BP2-BP7 has been discussed in Ref \cite{Abouabid:2022rnd}. 

It is worthy to point out that although the production rate of BP1 is large at the collision $\sqrt{s}=250$ GeV, the final states would be $ H^+H^- \to W^{+(*)} W^{-(*)} H^0 H^0$. Due to the small mass split between charged scalar boson and $H^0$ ($\Delta m = m_{H^\pm} -m_{H^0} = 0.7$ GeV), the charged W bosons must be off-shell and their visible decay products must be very soft and most of the invisible decay products can escape from the detectors, which poses a challenge to detect the signal events.

\begin{table}[!htb]
   {\scriptsize
    \renewcommand\arraystretch{1}
    \centering
    \begin{tabular}{|c|rrrrrrr|}
        \hline\hline
        BP                               & BP1    & BP2                 & BP3                  & BP4                  & BP5          & BP6         & BP7        \\
        \hline
        $m_{H^\pm}$ (GeV)                & 89.1   & 116.8               & 123.4                & 209.5                & 243.7        & 295.4       &  472.9         \\\hline
        $m_{H^0}$ (GeV)                  & 88.4   & 57.0                & 121.9                & 122.9                & 59.3         & 204.1        & 181.4        \\\hline
        $m_{A^0}$  (GeV)                 & 136.0   & 102.3               & 200.0                & 125.2                & 238.3        & 205.7        & 473.5        \\\hline
        $\mu_2^2$  (GeV$^2$)             & 7662.8 & 3159.5              & 14723.8              & 15037.5              & 3558.6       & 41195.9       &32220.8 
        \\\hline
        $\lambda_{L} (10^{-3})$          & 2.630      & 1.537               & 2.416                & 1.151                & -0.798       & 7.847      &11.819          \\\hline
        $\lambda_{S}$                    & 0.183     & 0.124               & 0.427                & 0.011                & 0.900        & 0.019        & 3.246        \\
        \hline
        $\Omega h^2 (\times 10^{-2})$    & 0.48   & 10.09               & 0.34                 & 0.20                 & 5.43         & 0.18          &0.028       \\\hline
        %$\Gamma_{H^\pm}$ (GeV)           & $1.08\times 10^{-13}$      & $5.87\times10^{-4}$ & $5.48\times10^{-12}$ & $6.77\times10^{-2}$  & $2.58$    & $2.18\times10^{-1}$ & $1.86\times 10^1$ \\\hline
         $\Gamma_{H^\pm}$ (GeV)           & $1.1\times 10^{-13}$      & $5.9\times10^{-4}$ & $5.5\cdot10^{-12}$ & $6.8\cdot10^{-2}$  & $2.6$       & $2.2\times10^{-1}$ &$1.9\times 10^1$ \\\hline
        $\Gamma_{A^0}$ (GeV)             & $5.8\times 10^{-4}$      & $1.1\times10^{-4}$ & $1.2\times10^{-2}$  & $5.2\times10^{-11}$ & $2.0$       & $8.8\times10^{-12}$ &$1.8\times 10^1$ \\\hline
        $Br(h^0 \to H^0H^0)$             & -      & $0.47\%$            & -                    & -                    & $0.10\%$     & -        &-            \\\hline
        $Br(A^0 \to W^{\pm(*)} H^{\mp})$ & $\sim71.0\%$      & -                   & $\sim 76.5\%$        & -         & -            & -    &$\sim 0 $\\ \hline
        $Br(A^0 \to Z^{(*)} H^0)$        & $\sim  29.0\%$   & $100\%$             & $\sim 23.5\%$        & $100\%$              & $100\%$      & $100\%$   &$\sim 100\%$           \\\hline
        $Br(H^\pm \to W^{\pm(*)} A^0)$   & -      & $\sim 0\%$          & -                    & $\sim 41.6\%$        & $\sim 0\%$   & $\sim 43.8\%$    & -    \\ \hline
        $Br(H^\pm \to W^{\pm(*)} H^0)$   & 100\%      & $\sim 100\%$        & $100\%$              & $\sim 58.4\%$        & $\sim 100\%$ & $\sim 56.2\%$     & $100\%$   \\
        \hline\hline
    \end{tabular}
    \caption{Benchmark points consistent with collider experiments  and dark matter constraints on the relic density are proposed. 
    Decay information of $H^0$, $A^0$ and $H^\pm$ are also given.}
    \label{tab:BPs}
    }
\end{table}

In Table \ref{Tab:XSBPs}, we presented the numerical results of LO/NLO cross sections and percentages of weak and QED corrections for the proposed benchmark points given in Table \ref{tab:BPs}. 

\begin{table}[!htb]
    \renewcommand\arraystretch{1.2}
    \centering
    \begin{tabular}{|c|crrrr|}
        \hline\hline
        \multicolumn{1}{|c|}{$\sqrt{s}$ (GeV) } & \multicolumn{1}{c}{BP}
                                                & \multicolumn{1}{c}{$\sigma^{0}$ (fb)}           & \multicolumn{1}{c}{$\Delta_{\mathrm{weak}}$(\%)}
                                                & \multicolumn{1}{c}{$\Delta_{\mathrm{QED}}$(\%)} & \multicolumn{1}{c|}{$\sigma^{\mathrm{NLO}}$ (fb)}                           \\
        \hline
        \multirow{3}*{250}
                                                & BP1                                             & 953.99                                            & 4.62   & -1.31     & 985.56      \\
                                                & BP2                                             & 731.24                                            & -7.06  & -1.13 & 671.42 \\
                                                & BP3                                             & 393.58                                            & -10.19 & -1.08 & 349.24 \\
        \hline
        \multirow{5}*{500}
                                                & BP1                                             & 387.93                                            & 3.83   & -0.56     & 400.61      \\
                                                & BP2                                             & 319.06                                            & 4.67   & -0.87 & 331.17 \\
                                                & BP3                                             & 305.45                                            & 6.17   & -0.94 & 321.42 \\
                                                & BP4                                             & 223.27                                            & -8.63  & -1.26 & 201.18 \\
                                                & BP5                                             & 129.57                                            & -5.40  & -1.08 & 121.18 \\
        \hline
        \multirow{7}*{1000}
                                                & BP1                                             & 128.81                                            & -3.22  & -0.07     & 124.74      \\
                                                & BP2                                             & 119.02                                            & -3.67  & -0.17 & 114.46 \\
                                                & BP3                                             & 116.65                                            & -1.93  & -0.22 & 114.14 \\
                                                & BP4                                             & 86.78                                             & -3.93  & -0.74 & 82.73  \\
                                                & BP5                                             & 77.13                                             & -0.37  & -0.92 & 76.13  \\
                                                & BP6                                             & 66.37                                             & -4.13  & -1.16 & 62.86  \\
                                                & BP7                                             & 43.09                                             & 3.38   & -1.11 & 44.06  \\
        \hline\hline
    \end{tabular}
    \caption{Weak corrections, QED corrections, the LO and full one-loop $\gamma\gamma \to H^\pm H^\mp$ cross sections of BPs are provided.}
    \label{Tab:XSBPs}
\end{table}

It is interesting to compare the results of the process $\gamma \gamma \to H^\pm H^\mp$ with those of the process $e^+ e^- \to  H^\pm H^\mp$ presented in \cite{Abouabid:2022rnd} and results in literatures.
\begin{itemize}
\item The cross section of $\gamma\gamma \to H^{\pm} H^{\mp}$ is around a few or tens times larger than that of $e^+ e^- \to H^\pm H^\mp$ when the mass of charged scalar boson and collision energy are fixed to be the same. For example,  for BP2 given in Table \ref{tab:BPs}, at the collision energy $\sqrt{s}=250$ GeV, the cross section of  $\gamma\gamma \to H^{\pm} H^{\mp}$  at the full NLO is 671 fb or so, while that of $e^+ e^- \to H^\pm H^\mp$ is only 22.5 fb. Such a feature can be attributed to the fact that the leading process of $\gamma\gamma \to H^{\pm} H^{\mp}$ includes both contact interaction and t/u-channels, while the leading process $e^+ e^- \to H^\pm H^\mp$ is s-channel dominant.
\item With the increase of collision energies from 250 GeV, to 500 GeV, and to 1 TeV, the total cross sections of $\gamma\gamma \to H^{\pm} H^{\mp}$ for BP2  are 671 fb, 331 fb, and 114fb.
But the total cross sections of $e^+ e^- \to H^{\pm} H^{\mp}$ for BP2 are 22 fb, 80 fb and 25 fb. With the same collision energy, $\gamma\gamma$ collision mode can have a larger cross section than that of $e^+ e^-$ collision mode, which is a well-known advantage of photon-photon collisions \cite{Boos:2000ki}.
\item In \cite{Zhu:1997es}, where the one-loop corrections of MSSM had been calculated,  the contributions of top's supersymmetric partners (top squarks) are found to modify the total cross section by a factor $-10\%$ or so. In \cite{Wang:2005pjs}, some benchmark points are studied and one-loop corrections are found to be between $-7\%--19\%$. Here in this work, we found that the trilinear Higgs couplings can affect the cross sections significantly, especially nearly the kinematic threshold regions when $\sqrt{s}$ is close to 1TeV and when the mass of charged scalar boson is heavy, which corresponds to a large trilinear Higgs coupling $\lambda_{h^0H^+ H^-}$ ($\sim -8$ or so), as shown in Fig. \ref{ggHH-degen} and Fig. \ref{fig:ggHH}. This large correction is due to those triangle and box diagrams which can contribute to terms proportional to $\lambda_{h^0H^+H^-}^2$.

\end{itemize}

\begin{table}[!htb]
    \renewcommand\arraystretch{1.2}
    \centering
    \begin{tabular}{|c|crrrr|}
        \hline\hline
        \multicolumn{1}{|c|}{$\sqrt{s}$ (GeV) } & \multicolumn{1}{c}{BP}
                                                & \multicolumn{1}{c}{$\sigma^{0}$ (fb)}           & \multicolumn{1}{c}{$\Delta_{\mathrm{weak}}$(\%)}
                                                & \multicolumn{1}{c}{$\Delta_{\mathrm{QED}}$(\%)} & \multicolumn{1}{c|}{$\sigma^{\mathrm{NLO}}$ (fb)}                           \\
        \hline
        \multirow{1}*{250}
                                                & BP1                                             & 0.76                                              & -9.2   & -1.10 & 0.68   \\
        \hline
        \multirow{4}*{500}
                                                & BP1                                             & 190.25                                            & 3.09    & -1.13 & 193.98 \\
                                                & BP2                                             & 103.31                                            & 1.19   & -1.24 & 103.25 \\
                                                & BP3                                             & 90.91                                             & 0.87   & -1.26 & 90.56  \\
                                                & BP4                                             & 0.06                                              & -13.19 & -1.08 & 0.05   \\
        \hline
        \multirow{6}*{1000}
                                                & BP1                                             & 117.04                                            & 1.98     & -0.93 & 118.26 \\
                                                & BP2                                             & 67.27                                             & 0.80   & -0.98 & 67.15  \\
                                                & BP3                                             & 59.67                                             & 1.50   & -0.98 & 59.98  \\
                                                & BP4                                             & 18.79                                             & -4.34  & -1.17 & 17.76  \\
                                                & BP5                                             & 13.31                                             & -2.64  & -1.22 & 12.79  \\
                                                & BP6                                             & 8.23                                              & -9.70  & -1.26 & 7.33   \\
        \hline\hline
    \end{tabular}
    \caption{Weak corrections, QED corrections, the LO and full one-loop $e^+ e^- \to \gamma\gamma \to H^\pm H^\mp$ cross sections of BPs are provided.}
    \label{Tab:partonXSBPs}
\end{table}

%In table \ref{Tab:partonXSBPs}, the cross section of physics processes $e^+ e^- \to \gamma\gamma \to H^\pm H^\mp$ are presented. 
%{\color{red}The cross section of the processes can be expressed as 
%\begin{equation}
%\cancel{\sigma=\int_{x_1+x_2 \geq 2 \frac{m_{H^\pm}}{\sqrt{s}}} f_{\gamma_1,e^+}(x_1) f_{\gamma_2,e^-}(x_2)  \hat{\sigma}(\gamma_1 \gamma_2 \to H^\pm H^\mp)}
%\end{equation}
%where the spectral functions $f_{\gamma_1, e^+}(x_1)$ and $f_{\gamma_2,e^-}(x_2)$ denote the distribution functions of back-scattered photons. And $x_1$ and $x_2$ represent the fractions of energy of photons with respect to the incoming positron/electron beams, which are always smaller than $1$.  The factor $f_{\gamma_1,e^+}(x_1) f_{\gamma_2,e^-}(x_2)$ can be regarded as the incoming fluxed of photons.}\\
In table \ref{Tab:partonXSBPs}, the cross sections of physics processes $e^+ e^- \to \gamma\gamma \to H^\pm H^\mp$ are presented. For $e^+e^-$ operation in photon collider mode we fold the partonic $\gamma\gamma \to H^\pm H^\mp$ cross section with the Compton backscattered photon spectra which are taken from
 CompAZ library \cite{ECFADESYPhotonColliderWorkingGroup:2001ikq,Ginzburg:1999wz,Ginzburg:1981vm,Ginzburg:1982yr,Telnov:1989sd}: %\cite{Zarnecki:2002qr,Ginzburg:1999wz,ECFADESYPhotonColliderWorkingGroup:2001ikq}:
\begin{equation}
	\sigma_{e^+e^-\to\gamma\gamma\to H^\pm H^\mp}(s)
	=\int_0^{x_{\max}} dx_1 \int_0^{x_{\max}} dx_2\;
	f_{\gamma/e}(x_1;\zeta)\, f_{\gamma/e}(x_2;\zeta)\;
	\hat\sigma_{\gamma\gamma\to H^\pm H^\mp }(x_1x_2 s)\,
	\Theta(x_1x_2 s - 4m_{H^\pm}^2),
\end{equation}
$x_1$ and $x_2$ represent the fractions of the incoming electron and positron beam energies carried by the back-scattered photons. Their values satisfy $0 < x_{1,2} < x_{\text{max}} < 1$, where  $x_{\max} = \zeta/(1+\zeta)$ and $\zeta \lesssim 4.8$ to avoid $e^+e^-$
pair creation in the conversion region \footnote{$\Theta$ is the Heaviside step function enforcing the product threshold.}.  The unpolarized Compton spectrum is: %\cite{Ginzburg:1981vm,Ginzburg:1982yr,Telnov:1989sd}
\begin{equation}
	f_{\gamma/e}(x;\zeta)=\frac{1}{\mathcal N(\zeta)}
	\left[
	1-x+\frac{1}{1-x}-\frac{4x}{\zeta(1-x)}+\frac{4x^2}{\zeta^2(1-x)^2}
	\right],\quad 0<x<x_{\max},
\end{equation}
\begin{equation}
	\mathcal N(\zeta)=\left(1-\frac{4}{\zeta}-\frac{8}{\zeta^2}\right)\ln(1+\zeta)
	+\frac12+\frac{8}{\zeta}-\frac{1}{2(1+\zeta)^2}.
\end{equation}

For the collision energy $\sqrt{s}=250$ GeV and BP1, the cross section is three orders smaller than the partonic one $\gamma\gamma \to H^\pm H^\mp$, which is due to the small incoming fluxes of photons. While for other BPs, like BP2 and BP3, although the collision energy is kinematically allowed in principle, but the cross sections are almost vanishing  due to the almost vanishing incoming fluxes of photons.

The situation becomes better when the collision energy is $\sqrt{s}=500$ GeV for BP1-BP4. For BP1, the cross sections of $e^+ e^- \to \gamma\gamma \to H^\pm H^\mp$ can be half of that of $\gamma\gamma \to H^\pm H^\mp$. Meanwhile, the cross section of  $e^+ e^- \to \gamma\gamma \to H^\pm H^\mp$ is also larger than that of $e^+ e^- \to H^\pm H^\mp$, due to the large incoming fluxes of photons convoluting with sizable cross sections. Similarly for BP2 and BP3, the cross sections of  $e^+ e^- \to \gamma\gamma \to H^\pm H^\mp$ is larger than $e^+ e^- \to H^\pm H^\mp$ by a percentage $29\%$ and $17\%$, respectively. While for BP4, although the cross section of $\gamma \gamma \to H^\pm H^\mp$ could be large, but the fluxes of incoming photons are too small to yield a large cross section. Such a fact also holds for BP5, where the cross section of this process can be safely negligible.

Like the case $\sqrt{s}=500$ GeV, when the collision energy is $\sqrt{s}=1$ TeV for BP1, the cross sections of $e^+ e^- \to \gamma\gamma \to H^\pm H^\mp$ can be of the same size as  that of $\gamma\gamma \to H^\pm H^\mp$, which is also much larger than that of $e^+ e^- \to H^\pm H^\mp$. For BP2 and BP3, the cross sections of  $e^+ e^- \to \gamma\gamma \to H^\pm H^\mp$ is a few times larger than that of $e^+ e^- \to H^\pm H^\mp$. While for BP4, the cross section of $e^+ e^- \to \gamma\gamma \to H^\pm H^\mp$ is almost the same as that of $e^+ e^- \to H^\pm H^\mp$. For BP5 and BP6, the contribution of charge scalar boson pair production from the process $e^+ e^- \to \gamma\gamma \to H^\pm H^\mp$ is considerable when compared with those of the process  $e^+ e^- \to H^\pm H^\mp$. For BP7, the cross section $e^+ e^- \to \gamma\gamma \to H^\pm H^\mp $ is negligible.

As illustrated in Figs.~4.1 and~4.2, the magnitude of the electroweak corrections to $\gamma\gamma \to H^\pm H^\mp$ can become very large in regions of parameter space where the absolute value of the trilinear coupling $\lambda_{h^{0}H^{+}H^{-}}$ is sizable, in particular near the kinematic threshold at high collision energies. It is worth emphasizing that this coupling also enters loop-induced Higgs observables, most notably the Higgs boson diphoton decay $h^{0}\to\gamma\gamma$ and the trilinear Higgs boson self-coupling $h^{0}h^{0}h^{0}$. Consequently, large values of $|\lambda_{h^{0}H^{+}H^{-}}|$ are subject to indirect constraints from current Higgs precision measurements at the LHC.
		
It is worth noting that in our analysis the IDM parameter space is scanned under all relevant theoretical and experimental constraints, as summarized in Section~2.2, and the benchmark points discussed in this work are representative of these allowed regions. While current data still permit sizable trilinear couplings $\lambda_{h^{0}H^{+}H^{-}}$  that can give rise to enhanced radiative effects, future Higgs precision measurements at the HL-LHC, especially in $h^{0}\to\gamma\gamma$ and double-Higgs production, are expected to further probe and constrain this part of the parameter space.

\section{Conclusions}
\label{sec:conclusions}
%%%%%%%%%%%%%%%%%%%%%%%%%%%%
In this work, we have carried out a comprehensive study of the  NLO electroweak corrections to the process $\gamma\gamma \to H^\pm H^\mp$ within the inert doublet model. The full set of one-loop contributions has been included, comprising vertex, box, self-energy, the corresponding counterterm, and real-emission diagrams. The Feynman gauge is used to generate Feynman amplitudes and then
they are computed using dimensional regularization. Ultraviolet divergences were consistently removed using an on-shell renormalization scheme, while infrared singularities were canceled by including both soft and hard photon radiation, in accordance with the KLN theorem. 

We have evaluated the impact of quantum effects on charged scalar pair production in photon-photon collisions. Our numerical results demonstrate that the $\gamma\gamma$ collision mode can exceed the $e^+e^-$ mode in production rate under comparable kinematic conditions, owing to the presence of contact and $t/u$-channel contributions. Near the production threshold, the NLO corrections can be particularly large, in some cases reaching or even surpassing 100\%, and thus play a decisive role in shaping the expected cross sections.

	A key insight from our analysis is the strong dependence of the production cross section and the size of quantum corrections on the collision energy and charged scalar mass. Increasing the collider energy from 250 GeV to 1 TeV results in a substantial reduction of the leading-order cross section; however, higher energies enable the exploration of parameter regions where the charged scalar is heavier, and the presence of large trilinear scalar couplings $|\lambda_{h^0H^+H^-}|$ can result
	in significant quantum effects, especially for the box diagrams that contain $\lambda_{h^0H^+H^-}^2$.
%	 large trilinear scalar couplings $|\lambda_{h^0H^+H^-}|$ induce sizable quantum effects.
	  Notably, near the 1 TeV threshold, the corrections can be very large: up to $+180\%$ in scenario I, $+120\%$ in scenario II, and $+60\%$ in scenario III, driven predominantly by triangle and box diagrams sensitive to the scalar self-couplings. These enhanced corrections underscore the importance of including NLO effects for precise phenomenological predictions.\\
The process $\gamma\gamma \to H^\pm H^\mp$ studied in this work provides information that is
	  	highly complementary to Higgs precision measurements at the LHC and its high-luminosity
	  	upgrade. While Higgs observables such as $h^{0}\to\gamma\gamma$ and di-Higgs production probe
	  	the effects of inert scalars indirectly through loop corrections, photon--photon collisions offer
	  	direct sensitivity to the charged scalar sector via pair production, supplemented by sizable
	  	radiative effects sensitive to the scalar potential parameters.
	  	In particular, potential deviations observed in Higgs boson loop-induced observables at the HL-LHC could be correlated with enhanced electroweak corrections in charged scalar pair
	  	production at photon colliders. Such correlations would provide a powerful discriminator
	  	between the IDM and other new-physics scenarios with extended scalar sectors,
	  	thereby strengthening the physics case for combining Higgs precision studies with
	  	$\gamma\gamma \to H^\pm H^\mp$ measurements at future lepton and photon colliders.
	  	
	Our detailed parameter space analysis reveals that the magnitude and sign of the radiative corrections are highly sensitive to the masses of the charged scalar and other IDM parameters, such as the scalar trilinear couplings. We find that the corrections can be either positive or negative, depending on the collider energy and specific benchmark points, emphasizing the need for careful theoretical modeling in experimental analyses.
	
	In conclusion, the $\gamma\gamma \to H^\pm H^\mp$ process serves as a highly promising channel for investigating the inert scalar sector, particularly at future high-energy photon colliders. The sizable quantum corrections elucidated in this work highlight their critical role in accurately predicting production rates and discerning the signatures of inert scalar particles. Our results offer valuable guidance for experimental design and analysis strategies aimed at probing the extended scalar sector beyond the SM, and underscore the necessity of incorporating full NLO corrections in theoretical predictions to maximize discovery potential.\\

\section*{Acknowledgements}
AA is supported by the Arab Fund for economic and social development.  B. Gong is supported by the Natural Science Foundation of China Nos.  11975242 and 12135013.  Q.S.Yan is supported by the Natural Science Foundation of China under the grant No. 11475180 and No. 11875260. We thank Jianxiong Wang for helpful discussions about FDC program. J. El Falaki would like to acknowledge B. Gong and Q.S. Yan for their warm hospitality
and financial support, which made it possible to complete part of this work.

%%%%%%%%%%%%%%%%%%%%%%%%%%%%

\appendix

\section{Feynman Diagrams}

This section presents the Feynman diagrams for  $\gamma \gamma \to H^\pm H^\mp$ at the one-loop level. The diagrams are categorized into vertex corrections, 
box contributions, self-energy and the corresponding counterterm corrections, and real emission diagrams.
%\begin{figure}[H]
%    \centering
%%    \includegraphics[width=1\textwidth]{Figures/vert1.png}
%%    \caption{Vertex feynman diagrams for $\gamma \gamma \to H^\pm H^\mp $}\label{fig:vert1}
%           \includegraphics[width=1\textwidth]{Figures/vert.pdf}
%\end{figure}
%
%\begin{figure}[H]
%    \centering
%    \includegraphics[width=1\textwidth]{Figures/vert2.png}
%    \caption{More vertex feynman diagrams for $\gamma \gamma \to H^\pm H^\mp $}\label{fig:vert2}
%\end{figure}

\begin{figure}[H]
    \centering
        \includegraphics[width=1\textwidth]{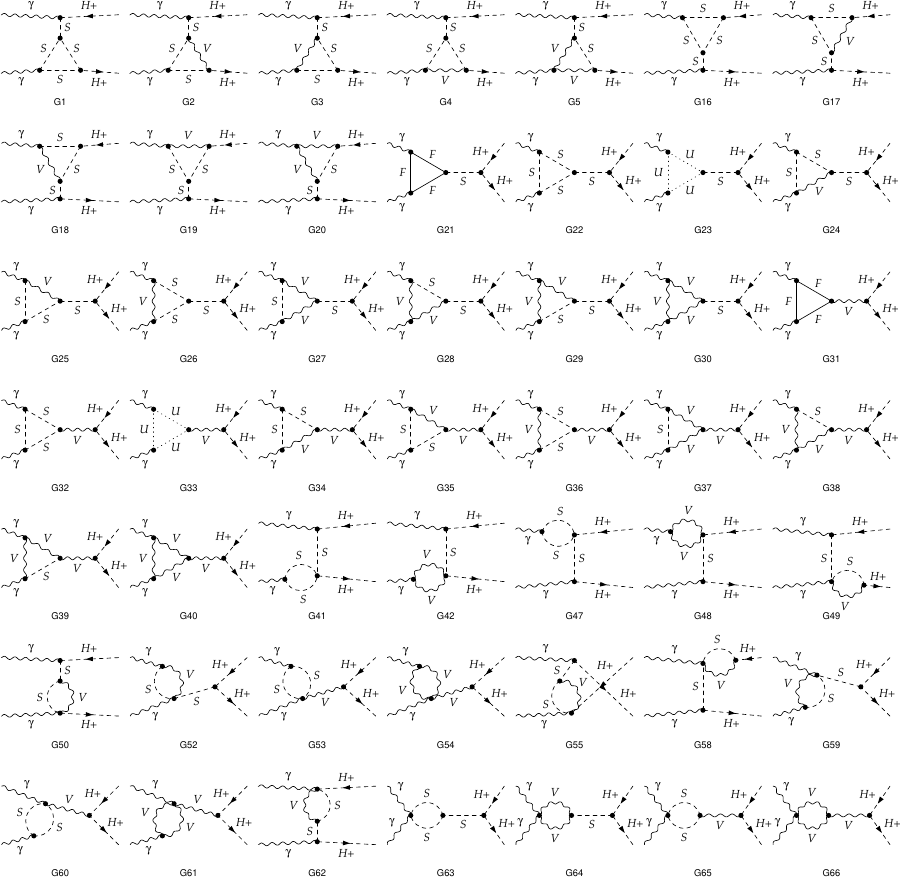}
    \caption{Vertex feynman diagrams for $\gamma \gamma \to H^\pm H^\mp $}\label{fig:vert1}
\end{figure}
Figure \ref{fig:vert1} displays the vertex correction diagrams involving scalar and gauge boson loops. 
These diagrams contribute to the renormalization of the $\gamma H^+ H^-$ and $Z^0H^+ H^-$ vertices.

\begin{figure}[H]
    \centering
    \includegraphics[width=1\textwidth]{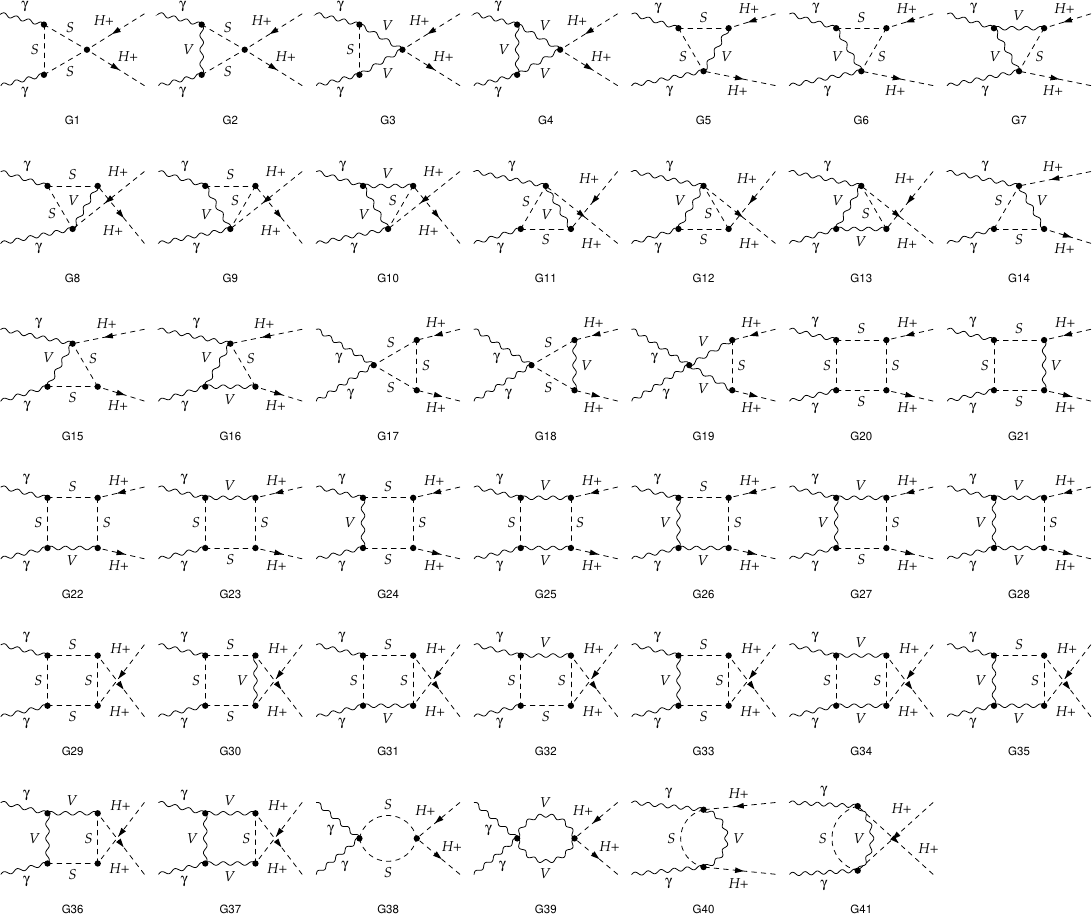}
    \caption{Box contributions to the loop level $\gamma \gamma \to H^\pm H^\mp $}\label{fig:box}
\end{figure}
Figure \ref{fig:box} illustrates the box diagrams with three different topologies:  $G_{1,...,19}$ , $G_{20,...,37}$ and $G_{38,...,41}$.
% which represent fourth-order corrections to the process. These diagrams are finite and gauge-invariant.

\begin{figure}[H]
    \centering
    \includegraphics[width=1\textwidth]{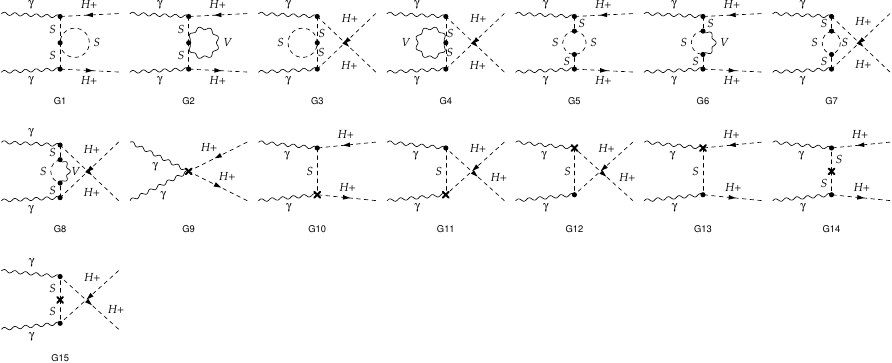}
    \caption{Self-energies and counterterms contributions Feynman diagrams for $\gamma \gamma \to H^\pm H^\mp $}
    \label{fig:self}
\end{figure}
Figure \ref{fig:self} presents the self-energy and the corresponding counterterm diagrams for $\gamma \gamma \to H^\pm H^\mp$. 
The self-energy corrections are needed both for the one-loop process as is depicted in $G_{1,...,8}$ and also for  the renormalisation of  the external charged boson.  
The set of counterterms ensure the cancellation of ultraviolet divergences.

\begin{figure}[H]
    \centering
    \includegraphics[width=1\textwidth]{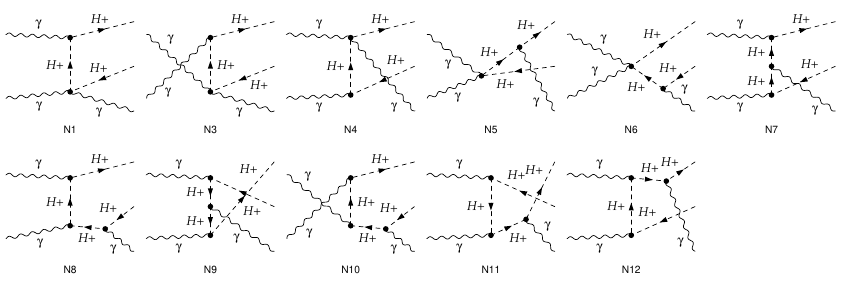}
    \caption{Real emission $\gamma \gamma \to H^\pm H^\mp \gamma$ contributions Feynman diagrams.}
    \label{fig:real}
\end{figure}
Figure \ref{fig:real} depicts the real emission diagrams for the process $\gamma \gamma \to H^\pm H^\mp \gamma$. These diagrams are necessary to cancel infrared divergences arising from virtual photon exchange in the one-loop amplitudes and wave function renormalisation of the charged scalar.

\section{\label{check:de}Check for independence on unphysical cut-off  $\Delta E$}

\begin{figure}[H]
	\includegraphics[width=\textwidth]{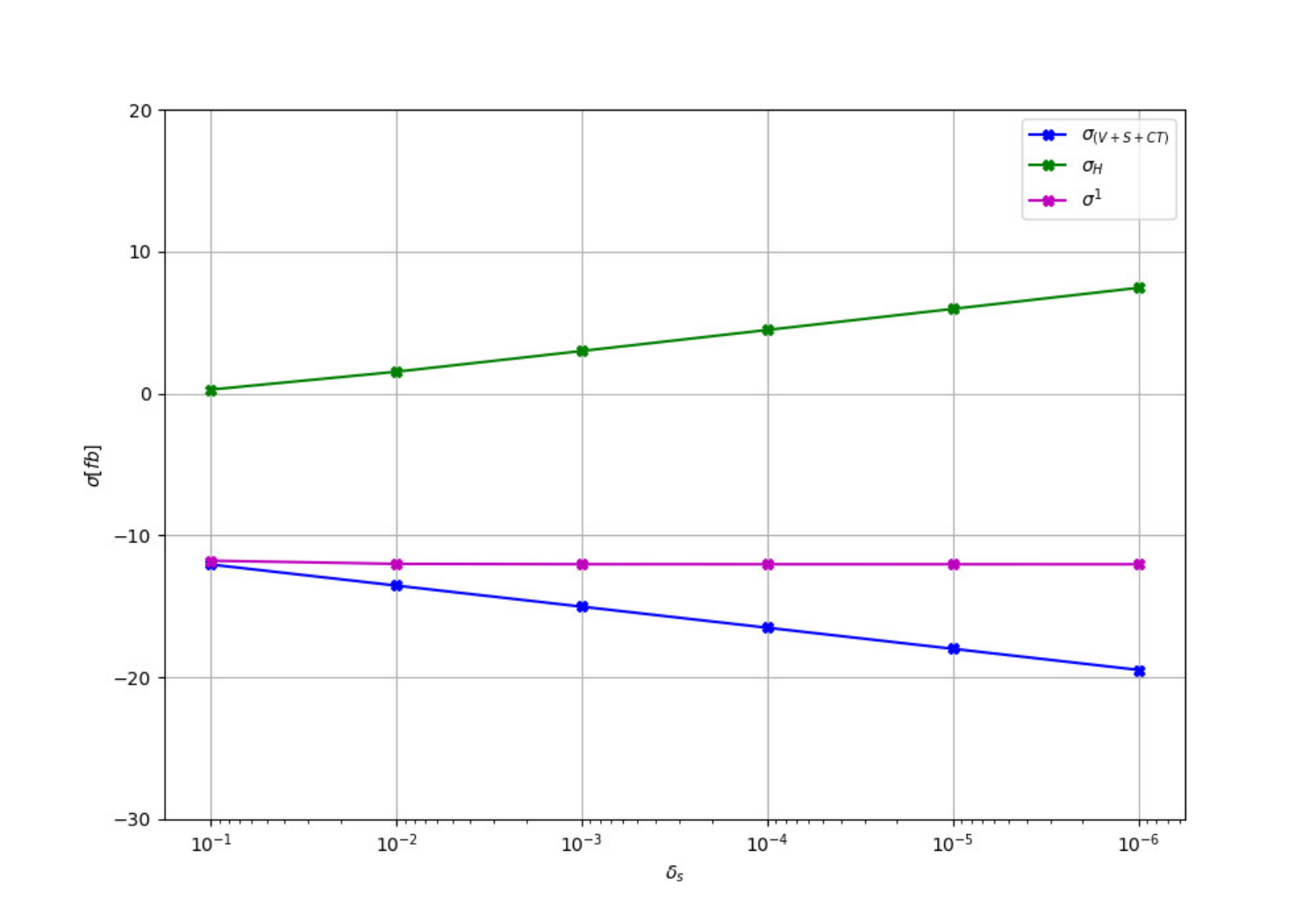}
	\caption{Dependence of the NLO contributions on the soft-photon cutoff 
		parameter $\delta_s$ for $\gamma\gamma \to H^\pm H^\mp$. }
		\label{QED-cuttoff}
\end{figure}

Fig.~\ref{QED-cuttoff} provides a graphical illustration of the infrared (IR) finiteness
check for $\gamma\gamma \to H^\pm H^\mp$ at NLO. It shows
how the different contributions depend on the soft-photon slicing parameter
$\delta_s$. Three curves are displayed: 
the hard real
emission contribution $\sigma_{\mathrm{H}}$ (computed with \texttt{FDC}),
the sum of remaining 
parts $\sigma_{\mathrm{V+S+CT}}$ (computed with \texttt{FormCalc}), 
and their sum $\sigma^{1}$, which represents the total NLO correction.
As $\delta_s$ is decreased over several orders of magnitude, 
$\sigma_{\mathrm{V+S+CT}}$ becomes more negative while $\sigma_{\mathrm{H}}$
increases accordingly, reflecting their individual dependence on the
arbitrary cutoff. In contrast, their sum $\sigma^{1}$ remains essentially
flat, demonstrating that the $\delta_s$ dependence cancels out, as required
by the KLN theorem. This behavior confirms the correct
implementation of the soft-photon subtraction and the consistency of our NLO
calculation.

\begin{table}[!h]
	\centering
	\begin{tabular}{c|c|c|c|}
	\hline
		\multicolumn{1}{|c|}{$\delta_s$} & $\sigma_{V+S+CT}(\mathrm{FormCalc})$ & $\sigma_H(\mathrm{FDC})$ & $\sigma^{(1)}$ \\ \hline
		\multicolumn{1}{|c|}{$10^{-1}$}  & -12.049(0)             & 0.272(0)     & -11.777   \\ \hline
		\multicolumn{1}{|c|}{$10^{-2}$}  & -13.533(0)             & 1.531(0)     & -12.002  \\ \hline
		\multicolumn{1}{|c|}{$10^{-3}$}  & -15.018(0)             & 2.993(1)     & -12.025   \\ \hline
		\multicolumn{1}{|c|}{$10^{-4}$}  & -16.502(0)             & 4.475(1)     & -12.027   \\ \hline
		\multicolumn{1}{|c|}{$10^{-5}$}  & -17.987(0)             & 5.959(2)     & -12.028   \\ \hline
		\multicolumn{1}{|c|}{$10^{-6}$}  & -19.472(0)             & 7.443(2)     & -12.029   \\ \hline
	\end{tabular}
	\caption{Check of the $\Delta E$ (soft cutoff) dependence for 
		$\gamma\gamma \to H^\pm H^\mp$, showing the numerical values of 
		$\sigma_{\mathrm{V+S+CT}}$, $\sigma_{\mathrm{H}}$, and their sum 
		$\sigma^{1}$ (in fb).}
		\label{tab-cutoff}
\end{table}

Table ~\ref{tab-cutoff}  complements Fig.~\ref{QED-cuttoff} by presenting the explicit numerical values
of $\sigma_{\mathrm{V+S+CT}}$, $\sigma_{\mathrm{H}}$, and their sum $\sigma^{1}$
for $\sqrt{s}=500$~GeV with 
$m_{H^\pm}=m_{H^0}=m_{A^0}=200$~GeV, $\lambda_2=2$, and $\mu_2^2=0$.
The results show that while the separate pieces $\sigma_{\mathrm{V+S+CT}}$ and
$\sigma_{\mathrm{H}}$ vary strongly with $\delta_s$, their sum stabilizes
around $-12.0$~fb across six orders of magnitude in $\delta_s$. The numerical
agreement at the $10^{-3}$~fb level demonstrates that the cancellation of the
cutoff dependence is achieved with high precision. Moreover, the fact that
the two components are obtained using independent codes (\texttt{FormCalc}
for $\sigma_{\mathrm{V+S+CT}}$ and \texttt{FDC} for $\sigma_{\mathrm{H}}$)
provides an additional robustness check of the calculation.\\

\newpage

\bibliographystyle{unsrt}
\bibliography{biblio}

\end{document}